\documentclass[aps,prd,11pt,superscriptaddress,notitlepage,longbibliography,nofootinbib,tightenlines,preprintnumbers]{revtex4-1}

\usepackage{booktabs}
\usepackage{amsmath,amssymb,amsfonts}
\usepackage{graphicx}
\usepackage{color}
\usepackage{tikz}
\usetikzlibrary{calc}
 \usetikzlibrary{decorations.text}
 \usetikzlibrary{decorations.pathreplacing}
 \usetikzlibrary{shapes}

\usepackage{dsfont}
\usepackage[pdftex]{hyperref}
\hypersetup{colorlinks=true, linkcolor=darkred, citecolor=blue, linktoc=page}
\definecolor{darkred}{rgb}{0.8,0.1,0.1}

\makeatletter\def\l@subsubsection#1#2{}%
\makeatother

\numberwithin{equation}{section}
\renewcommand\theequation{\arabic{section}.\arabic{equation}} 

\usepackage{subfigure}

\def\cA{{\cal A}}
\def\cB{{\cal B}}

\def\cG{{\cal G}}

\def\cN{{\cal N}}

\def\id{{\mathds{1}}}

\def\CC{\ensuremath{\mathds C}}
\def\RR{\ensuremath{\mathds R}}
\def\ZZ{\ensuremath{\mathds Z}}

\DeclareMathOperator{\vol}{vol}
\DeclareMathOperator{\Vol}{Vol}

\DeclareMathOperator{\Li}{Li}

\DeclareMathOperator{\sign}{sign}

\def\Im{\mathop{\rm Im}}

\def\Res{\mathop{\rm Res}}

\newcommand{\nocontentsline}[3]{}
\newcommand{\tocless}[2]{\bgroup\let\addcontentsline=\nocontentsline#1{#2}\egroup}

\begin{document}

\title{Surface defects in holographic 5d SCFTs}

\author{Michael Gutperle}
\email{gutperle@ucla.edu}  

\affiliation{Mani L. Bhaumik Institute for Theoretical Physics\\
	Department of Physics and Astronomy\\
	University of California, Los Angeles, CA 90095, USA
}

\author{Christoph F.~Uhlemann} 
\email{uhlemann@umich.edu}

\affiliation{Leinweber Center for Theoretical Physics, Department of Physics
	\\
	University of Michigan, Ann Arbor, MI 48109-1040, USA}

\preprint{LCTP-20-30}

\begin{abstract}
We use holography to study codimension-2 surface defects in 5d SCFTs engineered by $(p,q)$ 5-brane webs. 
The three-dimensional  defects are realized by D3-branes ending on the brane web.
We identify the holographic representation of the defects in Type IIB $AdS_6$ solutions as probe D3-branes, and study conformal and non-conformal defects which, respectively, preserve one half and one quarter of the supersymmetry. 
For a sample of 5d SCFTs, including the $T_N$ theories, we provide explicit solutions for conformal and non-conformal defects.
For the conformal defects we obtain their contribution to the free energy on $S^5$.
\end{abstract}

\maketitle
 
\tableofcontents

\setlength{\parskip}{1.5 pt}

\section{Introduction}

Defects are important objects  quantum field theory (QFT). They are sensitive to global aspects not captured by local operators, and can serve as non-local order parameters distinguishing different phases. A celebrated example is the area law for Wilson loops in confining gauge theories in four dimensions. Defects also lead to interesting mathematical structures such as generalized $q$-form symmetries \cite{Gaiotto:2014kfa} and higher group global symmetries \cite{Cordova:2018cvg}. 

In this paper we study codimension-$2$ surface defects in five-dimensional superconformal field theories (SCFTs).
We will focus on 5d SCFTs realized by $(p,q)$ 5-brane webs \cite{Aharony:1997ju,Aharony:1997bh} and study codimension-2 defects, which generalize surface defects in 4d QFTs.
Many 5d SCFTs can be understood as strongly-coupled UV fixed points of gauge theories \cite{Seiberg:1996bd,Intriligator:1997pq,Bhardwaj:2020gyu}, but the space of theories that can be realized by  5-brane webs and other string theory theory constructions is much broader.
Aspects of higher symmetries in 5d were discussed recently in \cite{Morrison:2020ool,Albertini:2020mdx}, and a classification program for defects in generic dimensions was initiated in \cite{Agmon:2020pde}.
Here we will use the holographic duals for 5d SCFTs engineered by $(p,q)$ 5-brane webs, constructed in \cite{DHoker:2016ujz,DHoker:2016ysh,DHoker:2017mds,DHoker:2017zwj}, as tool to study defects realized by D3-branes ending on the 5-brane webs \cite{Gaiotto:2014ina,Benvenuti:2016wet,Ashok:2017lko,Zenkevich:2017ylb,Nieri:2018ghd,Nieri:2018pev,Aprile:2018oau}.
We will study conformal and non-conformal defects, and obtain the defect contribution to the SCFT free energy on $S^5$ for conformal defects.
We hope this will be a starting point for fruitful interplay between AdS/CFT and field theory methods like supersymmetric localization in the study of these surface defects.

From a holographic perspective, defects with a small number of degrees of freedom compared to the degrees of freedom of the ambient SCFT are particularly accessible. They can be realized through probe branes \cite{Karch:2000gx}, whose backreaction can be neglected at leading order in the planar limit.\footnote{Examples of fully backreacted solutions describing superconformal defects can be found e.g.\ in \cite{DHoker:2007zhm,DHoker:2007mci,DHoker:2008lup,Drukker:2008wr}.}
A distinguished class of defects in conformal field theories are conformal defects, which preserve a conformal sub-algebra, $SO(p+1,2)$ for a $p+1$ dimensional defect, of the full conformal algebra of the ambient CFT. In superconformal field theories this may be enhanced to a superconformal sub-algebra. For conformal defects the dual probe branes wrap an $AdS_p$ subspace in the background geometry. The same probe brane embedding can then be used to describe planar defects in flat space and spherical defects in the ambient CFT on a sphere, depending on the coordinates chosen for the $AdS$ factor in the background geometry. More general defects can be obtained by using operators localized on the defect to trigger ``defect RG flows'', leading to a breaking of the defect conformal symmetry to the isometries preserved by the defect.

We will identify D3-brane embeddings describing superconformal defects in 5d SCFTs engineered by $(p,q)$ 5-brane webs. The D3-branes wrap $AdS_4$ in the $AdS_6$ part of the holographic duals and are localized at a distinguished point in the internal space. We will obtain the contribution of the defects to the SCFT free energy on $S^5$ from the brane on-shell action.
We will also study more general non-conformal defects, realizing RG flows triggered by defect-localized operators for planar surface defects. 
The D3-branes still wrap an $AdS_4$ in the background geometry, but the position in the internal space now changes along the radial coordinate, thus breaking the $AdS_4$ isometries corresponding to defect conformal transformations. 

The paper is organized as follows: In section \ref{sec2} we review the construction of 5d SCFTs with surface defects using 5-brane webs with D3-branes. 
In section \ref{sec3} we identify the holographic representation of the surface defects and find explicit solutions for supersymmetric conformal and non-conformal probe D3-brane embeddings. We also discuss the defect contribution to the free energy. We conclude in section \ref{sec4}.
The probe brane BPS conditions are derived in appendix \ref{app:D3-kappa}.

\section{D3-brane surface defects in 5-brane webs}\label{sec2}

We briefly review aspects of 5-brane webs \cite{Aharony:1997bh} that will be relevant and discuss surface defects realized by D3-branes, to guide the holographic discussion of surface defects in the next section.

Supersymmetric configurations of $(p,q)$ 5-branes, where we denote D5-branes by $(1,0)$ and NS5-branes by $(0,1)$, can be realized if all branes extend along the (01234) directions and are at an angle in the (56) plane such that $p\Delta x_5=q\Delta x_6$ (with the axion-dilaton scalar $\tau=i$).
General planar junctions of $(p,q)$ 5-branes at a point in the (56)-plane then define 5d SCFTs with 8 Poincar\'e and 8 superconformal supersymmetries. The $SU(2)$ R-symmetry is realized by rotations in the remaining (789) directions.

The Coulomb branch of the SCFT is realized by resolving the 5-brane junction at a point into a 5-brane web, while mass deformations are realized by moving the external 5-branes relative to each other. This can often be used to obtain an effective description of the SCFT as a gauge theory. Our focus will be on the fixed-point SCFTs, described by 5-brane junctions at a point, but to highlight the features of the SCFTs we will nevertheless show resolved webs in the figures (e.g.\ in figs.~\ref{fig:TN-web}, \ref{fig:YN-web}, \ref{fig:plus-web}, which show the (56) plane).
\begin{center}
	\def\arraystretch{1}
	\begin{tabular}{cccccc|cc|ccc}
		\toprule
		& \ 0 \ & \ 1 \ & \ 2 \ & \ 3 \ & \ 4 \ & \ 5 \ & \ 6 \ & \ 7 \ & \ 8 \ & \ 9 \ \\
		\hline
		D5-brane & $\times$ & $\times$ & $\times$ & $\times$ & $\times$ & $\times$ & \\
		NS5-brane & $\times$ & $\times$ & $\times$ & $\times$ & $\times$ &  & $\times$ \\
		$(p,q)$ 5-brane & $\times$ & $\times$ & $\times$ & $\times$ & $\times$ & \multicolumn{2}{c|}{angle}& \\
		$[p,q]$ 7-brane & $\times$ & $\times$ & $\times$ & $\times$ & $\times$ & & & $\times$ & $\times$ & $\times$\\
		\hline
		D3-brane & $\times$ & $\times$ & $\times$ &  & & & & $\times$ & & \\
		\bottomrule
	\end{tabular}
\end{center}
More general SCFTs can be realized by including in addition 7-branes extending in the (01234) and (789) directions \cite{DeWolfe:1999hj}. An example is shown in fig.~\ref{fig:I-web} below, and the brane orientations are summarized in the table above.

Our main interest are codimension-2 surface defects in the 5d SCFTs engineered by 5-brane junctions, which can be realized by adding D3-branes that end on the 5-brane web.
The D3-branes extend along 3 of the 5 field theory directions, say (012), and one of the directions which none of the 5-branes extend into, say (7). Various examples of such configurations were studied e.g.\ in \cite{Aharony:1997ju,Gaiotto:2014ina,Benvenuti:2016wet,Ashok:2017lko,Zenkevich:2017ylb,Nieri:2018ghd,Nieri:2018pev,Aprile:2018oau}.
From the field theory perspective these defects can be understood as extra degrees of freedom coupling to the 5d SCFT at the intersection point of the 5-brane junction and the D3-brane. They can also be obtained from RG flows with vortex configurations on the Higgs branch~\cite{Gaiotto:2014ina}.

\begin{figure}
\subfigure[][]{\label{fig:one-side-conf}
		\begin{tikzpicture}[x={(-0.866cm,0.5cm)}, y={(0.866cm,0.5cm)}, z={(0cm,1cm)}, scale=0.9]
		\foreach \i in {-0.2,-0.1,0,0.1,0.2}{
			\draw[thick] (\i,-1.8,0) -- (\i,1.8,0);}
		\foreach \j in {-0.1,0,0.1}{
			\draw[thick] (2.2,\j,0) -- (-2.2,\j,0);
		}
		\draw[very thick,blue] (0,0,0) -- (0,0,1.8);
		
		\draw[->] (-2,-2) -- (-2+.5,-2) node [below,font=\footnotesize] {5};
		\draw[->] (-2,-2) -- (-2,-2+.5) node [below,font=\footnotesize] {6};
		\draw[->] (-2,-2) -- (-2+.5,-2+.5) node [right,font=\footnotesize] {7};
	\end{tikzpicture}
}
\subfigure[][]{\label{fig:one-side-non-conf}
	\begin{tikzpicture}[x={(-0.866cm,0.5cm)}, y={(0.866cm,0.5cm)}, z={(0cm,1cm)}, scale=0.9]
	\foreach \i in {-0.2,-0.1,0,0.1,0.2}{
		\draw[thick] (\i,-1.8,0) -- (\i,1.8,0);}
	\foreach \j in {-0.1,0,0.1}{
		\draw[thick] (2.2,\j,0) -- (-2.2,\j,0);
	}
	\draw[very thick,blue] (0,-1,0) -- (0,-1,1.8);
	
	\draw[->,white] (-2,-2) -- (-2+.5,-2) node [below,font=\footnotesize] {5};
	\draw[->,white] (-2,-2) -- (-2,-2+.5) node [below,font=\footnotesize] {6};
	\draw[->,white] (-2,-2) -- (-2+.5,-2+.5) node [right,font=\footnotesize] {7};
\end{tikzpicture}
}
\subfigure[][]{\label{fig:two-side-conf}
		\begin{tikzpicture}[x={(-0.866cm,0.5cm)}, y={(0.866cm,0.5cm)}, z={(0cm,1cm)}, scale=0.9]
	\foreach \i in {-0.2,-0.1,0,0.1,0.2}{
		\draw[thick] (\i,-1.8,0) -- (\i,1.8,0);}
	\foreach \j in {-0.1,0,0.1}{
		\draw[thick] (2.2,\j,0) -- (-2.2,\j,0);
	}
	\draw[very thick,blue] (0,0,0) -- (0,0,1.8);
	\draw[very thick,dashed,blue] (0,0,0) -- (0,0,-1.8);
	
	\draw[->,white] (-2,-2) -- (-2+.5,-2) node [below,font=\footnotesize] {5};
	\draw[->,white] (-2,-2) -- (-2,-2+.5) node [below,font=\footnotesize] {6};
\end{tikzpicture}
}
\subfigure[][]{\label{fig:two-side-non-conf}
	\begin{tikzpicture}[x={(-0.866cm,0.5cm)}, y={(0.866cm,0.5cm)}, z={(0cm,1cm)}, scale=0.9]
	\foreach \i in {-0.2,-0.1,0,0.1,0.2}{
		\draw[thick] (\i,-1.8,0) -- (\i,1.8,0);}
	\foreach \j in {-0.1,0,0.1}{
		\draw[thick] (2.2,\j,0) -- (-2.2,\j,0);
	}
	\draw[very thick,blue] (-0.7,0.7,0) -- (-0.7,0.7,1.8);
	\draw[very thick,dashed,blue] (-0.7,0.7,0) -- (-0.7,0.7,-1.8);
	
	\draw[->,white] (-2,-2) -- (-2+.5,-2) node [below,font=\footnotesize] {5};
	\draw[->,white] (-2,-2) -- (-2,-2+.5) node [below,font=\footnotesize] {6};
\end{tikzpicture}
}
	\caption{From left to right: one-sided conformal defect, one-sided non-conformal defect, two-sided conformal defect, two-sided non-conformal defect. The black lines represent 5-branes in the (56) plane; the 5-brane junction has been resolved slightly for illustrative purposes.
	The vertical blue line represents a D3-brane, with the part above/below the (56) plane shown as solid/dashed line.\label{fig:defect-D3}}
\end{figure}
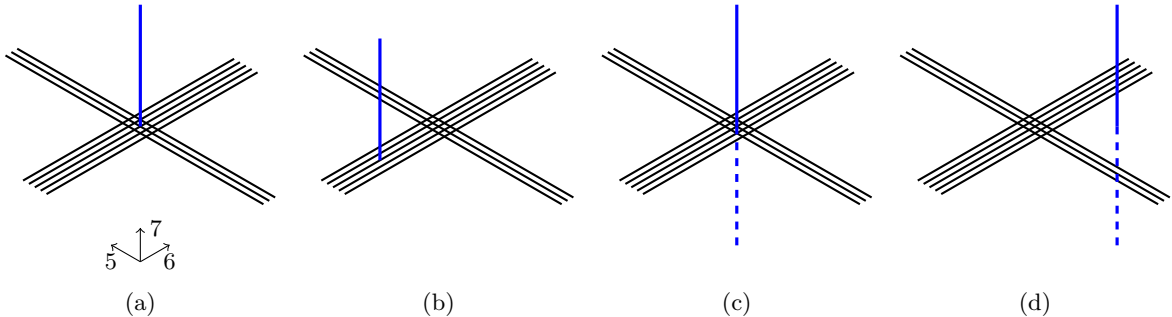

In this work we use AdS/CFT to study supersymmetric  defects realized by D3-branes in 5d SCFTs. That is, the ambient QFT is at the conformal fixed point.
If the position of the D3-brane in the (89) plane is fixed, the D3-brane preserves a $U(1)$ subgroup of the R-symmetry corresponding to rotation in the (89) plane.
All D3-branes studied below preserve this $U(1)$ symmetry.
We will distinguish between one-sided and two-sided defects: 
For one-sided defects the D3-brane ends on the 5-brane junction.
For a conformal defect the D3-brane is located in the (56) plane at the point of the 5-brane junction (fig.~\ref{fig:one-side-conf}). 
Non-conformal defects can be realized by D3-branes ending on one of the external 5-branes away from the junction point, with the separation from the junction point introducing a mass parameter, see fig.~\ref{fig:one-side-non-conf}.
For two-sided conformal defects, two D3-branes are joined at the point of the 5-brane junction from opposite sides (fig.~\ref{fig:two-side-conf}).
More general mass deformations can in that case be obtained by arbitrarily moving the complete D3-brane away from the junction point in the (56) plane (fig.~\ref{fig:two-side-non-conf}).
These features will be reflected in the discussion of probe D3-branes in the next section.

\section{D3-brane surface defects in \texorpdfstring{$AdS_6\times S^2\times\Sigma$}{Type IIB AdS6 solutions}} \label{sec3}

We briefly review the Type IIB supergravity solutions describing $(p,q)$ 5-brane junctions, constructed in \cite{DHoker:2016ujz,DHoker:2016ysh,DHoker:2017mds}, and then discuss D3-brane embeddings describing surface defects.

The $AdS_6$ solutions are defined in terms of locally holomorphic functions $\cA_\pm$ on a Riemann surface $\Sigma$ with boundary. The geometry is a warped product of $AdS_6$ and $S^2$ over $\Sigma$, with the $S^2$ collapsing on the boundary of $\Sigma$.
The Einstein-frame metric, complex two-form $C_{(2)}$, and axion-dilaton scalar $B=(1+i\tau)/(1-i\tau)$ are given by
\begin{align}\label{eqn:ansatz}
	ds^2 &= f_6^2 \, ds^2 _{\mathrm{AdS}_6} + f_2^2 \, ds^2 _{\mathrm{S}^2} 
	+ 4\rho^2\, |dw|^2~,
\qquad\quad
	B =\frac{\partial_w \cA_+ \,  \partial_{\bar w} \cG - R \, \partial_{\bar w} \bar \cA_-   \partial_w \cG}{
		R \, \partial_{\bar w}  \bar \cA_+ \partial_w \cG - \partial_w \cA_- \partial_{\bar w}  \cG}~,
\nonumber\\
	C_{(2)}&=\frac{2i}{3}\left(
\frac{\partial_{\bar w}\cG\partial_w\cA_++\partial_w \cG \partial_{\bar w}\bar\cA_-}{3\kappa^{2}T^2} - \bar{\mathcal{A}}_{-} - \mathcal{A}_{+}  \right)\vol_{S^2}~,
\end{align}
where $w$ is a complex coordinate on $\Sigma$ and $ds^2_{AdS_6}$ and $ds^2_{S^2}$ are the line elements for unit-radius $AdS_6$ and $S^2$, respectively.
The metric functions read
\begin{align}\label{eq:metric-functions}
	f_6^2&=\sqrt{6\cG T}~, & f_2^2&=\frac{1}{9}\sqrt{6\cG}\,T ^{-\tfrac{3}{2}}~, & \rho^2&=\frac{\kappa^2}{\sqrt{6\cG}} T^{\tfrac{1}{2}}~,
\end{align}
and we have 
\begin{align}\label{eq:kappa2-G}
	\kappa^2&=-|\partial_w \cA_+|^2+|\partial_w \cA_-|^2~,
	&
	\partial_w\cB&=\cA_+\partial_w \cA_- - \cA_-\partial_w\cA_+~,
	\nonumber\\
	\cG&=|\cA_+|^2-|\cA_-|^2+\cB+\bar{\cB}~,
	&
	T^2&=\left(\frac{1+R}{1-R}\right)^2=1+\frac{2|\partial_w\cG|^2}{3\kappa^2 \, \cG }~.
\end{align}
Explicit expressions for the functions $\cA_\pm$ will be given below. The differentials $\partial\cA_\pm$ generally have poles at isolated points $r_\ell$ on the boundary of $\Sigma$, at which 5-branes with charge $(p_\ell,q_\ell)$ emerge, with $(p_\ell,q_\ell)$ given in terms of the residues of $\partial\cA_\pm$ by 
\begin{align}\label{eq:charges}
	\Res_{w=r_\ell} \partial_w\cA_\pm & = \frac{3}{4}\alpha^\prime (\pm q_\ell +i p_\ell)~.
\end{align}
This allows to identify the associated 5-brane junction. Solutions for 5-brane junctions with 7-branes in addition have punctures with $SL(2,\RR)$ monodromy in the interior of $\Sigma$ \cite{DHoker:2017zwj}.

\subsection{D3-brane embeddings}\label{sec:D3-embeddings}

We now discuss probe D3-branes embedded into the Type IIB $AdS_6$ solutions, of a form appropriate to describe surface defects. 
We choose $AdS_6$ coordinates such that
\begin{align}
	ds^2_{AdS_6}&=dr^2+ e^{2r}\left(dx^\mu dx_\mu + dy_1^2+dy_2^2\right)~,
\end{align}
where $\mu=0,1,2$ are the field theory directions in which the defect should extend and $(y_1,y_2)$ is the plane transverse to the surface defect.
From the field theory perspective there is a geometric $SO(2)$ symmetry corresponding to rotations in the plane transverse to the surface defect.
This symmetry should be preserved by the D3-brane, which thus has to be located at a fixed position in the $(y_1,y_2)$ plane.
The worldvolume coordinates can be chosen as $(r,x^\mu)$.

The isometries of the $S^2$ in the background solution represent  the $SU(2)$ R-symmetry of the 5d SCFT, and we will consider only embeddings which preserve a $U(1)$ subgroup of the R-symmetry. This forces the position of the D3-brane on the $S^2$ to be constant across the embedding.
For a defect preserving 3d Poincar\'e symmetry, the only remaining freedom is then the position on $\Sigma$, and the entire embedding is characterized by one complex embedding function $w(r)$.
The induced metric for such an embedding is 
\begin{align}
	g_{\alpha\beta}d\xi^\alpha d\xi^\beta&=\left(f_6^2+4\rho^2 \partial_r w \partial_r \bar w\right)dr^2+e^{2r}dx^\mu dx_\mu~,
\end{align}
and the action for the D3-brane becomes
\begin{align}\label{d3act}
	S_{\rm D3}&=T_{\rm D3}\int dr\,d^3x\, f_6^3 e^{3r}\sqrt{f_6^2+4\rho^2 |w^\prime|^2}~,
	&
	T_{\rm D3}^{-1}&=(2\pi)^{3}{\alpha^\prime}^{2}~.
\end{align}
The equation of motion resulting from this action is a complex second-order non-linear ordinary differential equation for $w(r)$.
However, we are interested in supersymmetric defects and the BPS condition is more tractable.
The derivation is spelled out in app.~\ref{app:D3-kappa}, and results in the condition
\begin{align}\label{eq:BPS}
	\kappa^2 w^\prime &=\partial_{\bar w}\cG~.
\end{align}
As also shown in the appendix, this condition implies the equation of motion following from (\ref{d3act}).
As appropriate for a D3-brane, the BPS equation is invariant under Type IIB $SL(2,\RR)$ transformations, since $\kappa^2$ and $\cG$ are separately invariant.
The solutions flow along the gradient of $\cG$ in $\Sigma$, as dictated by (\ref{eq:BPS}), and generic embeddings of this form preserve 3d $\cN=2$ supersymmetry.

Embeddings preserving defect conformal symmetry, i.e.\ the $SO(2,3)$ isometries of the $AdS_4$ parametrized by $(r,x^\mu)$,
have $w^\prime=0$. The induced metric on the D3-brane is that of $AdS_4$ and the position on $\Sigma$ is a constant.
The BPS condition (\ref{eq:BPS}) reduces to 
\begin{align}\label{eq:BPS-massless}
	\partial_{\bar w}\cG&=0~.
\end{align}
The D3-brane thus is at an extremal point of $\cG$, which we denote as $w=w_c$. 
Such points are also extrema of $f_6$ (and $f_2$), so that the D3-brane extremizes its action.\footnote{For the solutions discussed below $\cG$ has a unique maximum in $\Sigma$. We are not aware of solutions with more than one maximum.}
As shown in appendix~\ref{app:D3-kappa} the embeddings preserve eight of the sixteen supersymmetries of the background, and realize the sub-superalgebra $C(3)$ of $F(4)$ \cite{Frappat:1996pb} (see also table 1 of \cite{Gutperle:2017nwo}).
The on-shell action for these embeddings can be evaluated using that $\partial_w\cG=0$ implies $T=1$, such that $f_6^4=6\cG$.
This leads to
\begin{align}\label{eq:D3-conf-action}
	S_{\rm D3}&=6T_{\rm D3}\Vol_{AdS_4}\cG|_{w=w_c}~.
\end{align}
So far we have worked with Poincar\'e $AdS_6$. However, for the embeddings with defect conformal symmetry one can do a conformal transformation to global $AdS_6$ coordinates, to describe an $S^3$ defect in the 5d SCFT on $S^5$. 
The renormalized volume of global $AdS_4$ is given by $\Vol_{AdS_4}=\frac{2}{3}\Vol_{S^3}$ with $\Vol_{S^3}=2\pi^2$, and with these values the on-shell action in (\ref{eq:D3-conf-action}) yields the contribution of the defect described by the D3-brane to the free energy of the SCFT on $S^5$.

An interpretation for the non-conformal embeddings described by (\ref{eq:BPS}) can be obtained by analyzing the behavior near the conformal boundary of $AdS_6$.
With a Fefferman-Graham radial coordinate $z=-\ln r$, such that the boundary is at $z=0$, the BPS equation becomes $-\kappa^2 z \partial_z w =\partial_{\bar w}\cG$.
At the boundary of $AdS_6$ the embedding approaches that of a conformal defect.
The mass of fluctuations in the real and imaginary parts of $w$ can be obtained by expanding (\ref{eq:BPS}) around the conformal embedding, using that (\ref{eq:BPS-massless}) implies $\cA_+-\bar\cA_-=0$ (since $\kappa^2(\cA_+-\bar\cA_-)=\partial_{\bar w}\bar\cA_-\partial_w\cG-\partial_{w}\cA_+\partial_{\bar w}\cG$ and $\kappa^2>0$ in the interior of $\Sigma$). This shows that the leading behavior of both fluctuations near the conformal boundary is linear in $z$.
Fluctuations in the real and imaginary parts of $w$ thus correspond to a pair of defect operators with scaling dimensions $\Delta=2$ in standard quantization and $\Delta=1$ in alternative quantization. The non-conformal embeddings we will discuss below describe defect RG flows triggered by combinations of these two relevant deformations.

The general BPS equation  (\ref{eq:BPS}) can be integrated to obtain an implicit expression for the embedding function $w(r)$.
To that end, we note that (\ref{eq:BPS}) implies
\begin{align}
	\frac{d}{dr}\left[e^r \left(\cA_+(w(r))-\overline{\cA_-(w(r))}\right)\right]&=0~.
\end{align}
Integrating both sides leads to
\begin{align}\label{eq:BPS-2}
 \cA_+(w(r))-\overline{\cA_-(w(r))}&=me^{-r}~,
\end{align}
where $m$ is a complex parameter encoding the mass parameters associated with the two relevant deformations discussed above.
The conformal embedding corresponds to $m=0$.
The Type IIB $SL(2,\RR)$ transformations are induced by $SU(1,1)\otimes \CC$ transformations of $\cA_\pm$, spelled out in (5.12) of \cite{DHoker:2016ujz}.
Under these transformations the combination $\cA_+-\overline{\cA_-}$  transforms by an overall factor, such that $SL(2,\RR)$ transformations of the condition in (\ref{eq:BPS-2}) only transform the mass parameter.

\subsection{\texorpdfstring{$T_N$}{T[N]} and \texorpdfstring{$Y_N$}{Y[N]} theories} \label{tytheories}

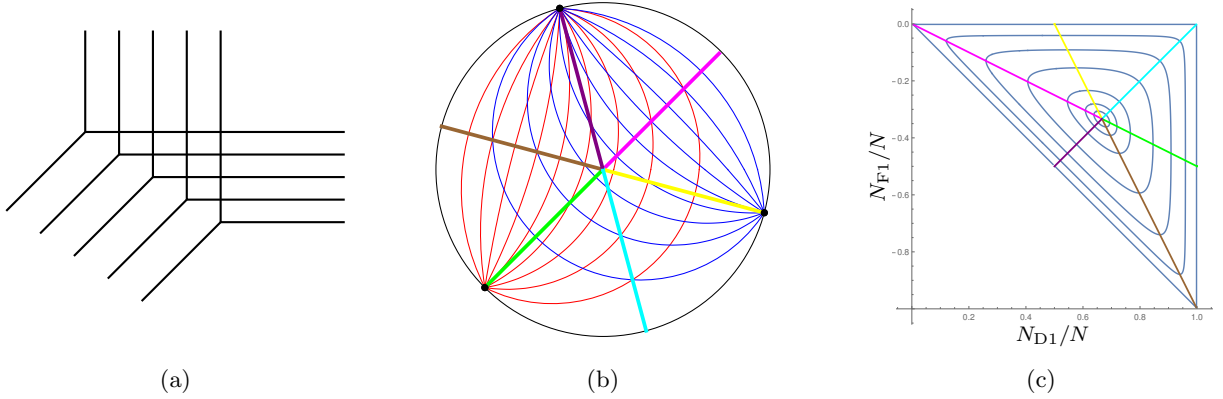
\begin{figure}
	\subfigure[][]{\label{fig:TN-web}
		\begin{tikzpicture}[xscale=-0.6,yscale=-0.6]
			\draw[thick] (-4,0.75) -- (-0.5,0.75) -- (-0.5,-3);
			\draw[thick] (-0.5,0.75) -- +(1.75,1.75);
			
			\draw[thick] (-4,0.25) -- (0.25,0.25) -- (0.25,-3);
			\draw[thick] (0.25,0.25) -- +(1.75,1.75);
			
			\draw[thick] (-4,-0.25) -- (1.0,-0.25) -- (1.0,-3);
			\draw[thick] (1.0,-0.25) -- +(1.75,1.75);
			
			\draw[thick] (-4,-0.75) -- (1.75,-0.75) -- (1.75,-3);
			\draw[thick] (1.75,-0.75) -- +(1.75,1.75);
			
			\draw[thick] (-4,1.25) -- (-1.25,1.25) -- (-1.25,-3);
			\draw[thick] (-1.25,1.25) -- +(1.75,1.75);
			
			\node at (0,4) {};
		\end{tikzpicture}
	}	\hskip 6mm
	\subfigure[][]{\label{fig:TN-disc}
		\includegraphics[width=0.3\linewidth]{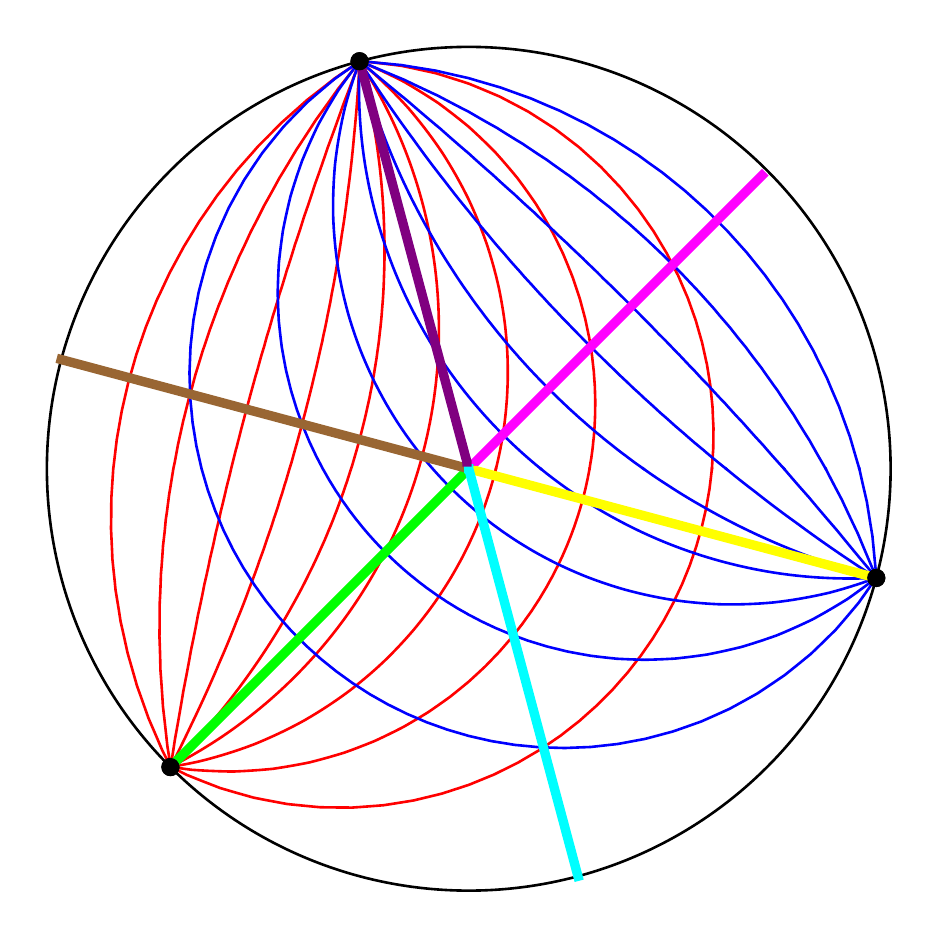}
	}	\hskip 6mm
	\subfigure[][]{\label{fig:TN-charges}
		\begin{tikzpicture}
		\node at (0,0) {\includegraphics[width=0.26\linewidth]{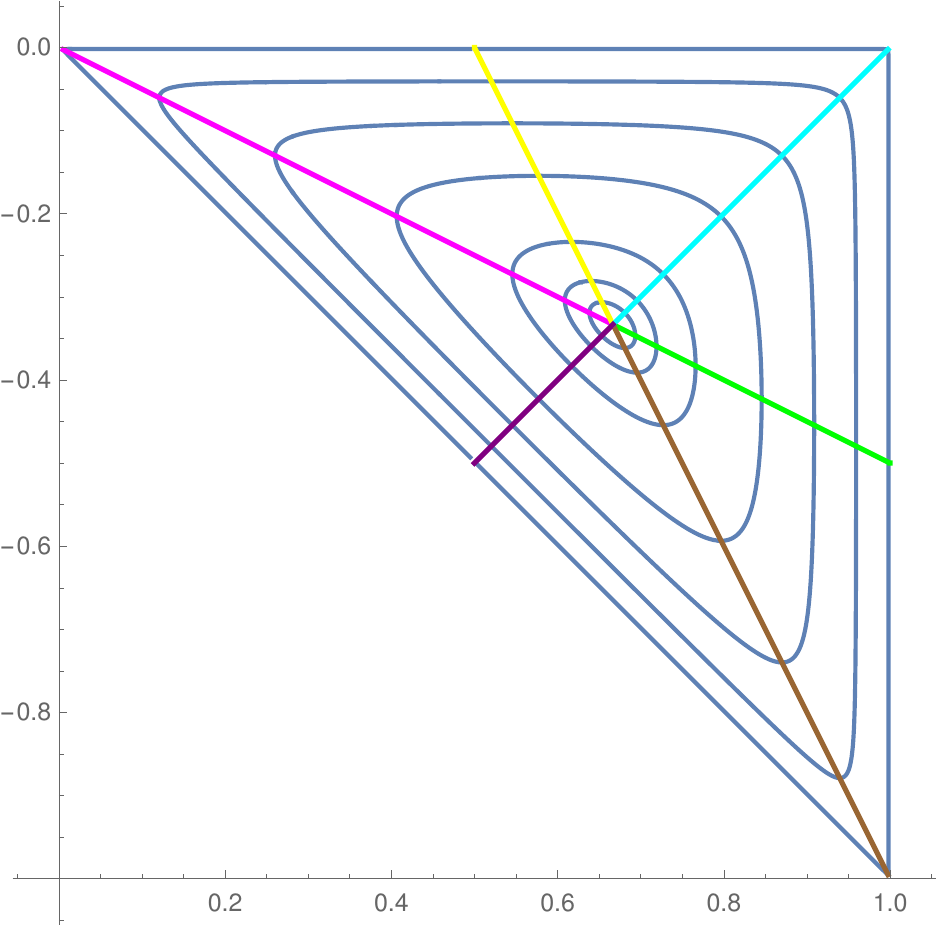}};
		\node[rotate=90] at (-2.3,0) {\scriptsize $N_{\rm F1}/N$};
		\node at (0,-2.25) {\scriptsize $N_{\rm D1}/N$};
		\end{tikzpicture}
	}
	\caption{Left: brane web for the 5d $T_N$ theory. Center: Supergravity solution represented on the disc. The (red,blue) curves show lines of constant $(N_{\rm F1},N_{\rm D1})$. Right: The blue curves show the $(N_{\rm D1},N_{\rm F1})$ charges along curves of constant radius on the disc, the edges of the triangle correspond to the poles.
		The D3-brane embeddings are: $w_1$ with positive/negative mass in magenta/green, $w_2$ with positive/negative mass in brown/yellow and $w_3$ with positive/negative mass in cyan/purple.\label{fig:TN}}
\end{figure}

\begin{figure}
	\subfigure[][]{\label{fig:YN-web}
\begin{tikzpicture}[scale=0.66]
	\draw[thick] (-0.25,-2.5) -- (-0.25,0.75) -- (0.25,0.75) -- (0.25,-2.5);
	\draw[thick] (-0.25,0.75) -- +(-1.5,1.5);
	\draw[thick] (0.25,0.75) -- +(1.5,1.5);
	\draw[thick] (-0.75,-2.5) -- (-0.75,0.25) -- (0.75,0.25) -- (0.75,-2.5);
	\draw[thick] (-0.75,0.25) -- +(-1.5,1.5);
	\draw[thick] (0.75,0.25) -- +(1.5,1.5);
	\draw[thick] (-1.25,-2.5) -- (-1.25,-0.25) -- (1.25,-0.25) -- (1.25,-2.5);
	\draw[thick] (-1.25,-0.25) -- +(-1.5,1.5);
	\draw[thick] (1.25,-0.25) -- +(1.5,1.5);
	\draw[thick] (-1.75,-2.5) -- (-1.75,-0.75) -- (1.75,-0.75) -- (1.75,-2.5);
	\draw[thick] (-1.75,-0.75) -- +(-1.5,1.5);
	\draw[thick] (1.75,-0.75) -- +(1.5,1.5);
	
	\node at (0,-3.3) {};
\end{tikzpicture}
	}	\hskip 5mm
	\subfigure[][]{\label{fig:YN-disc}
		\includegraphics[width=0.28\linewidth]{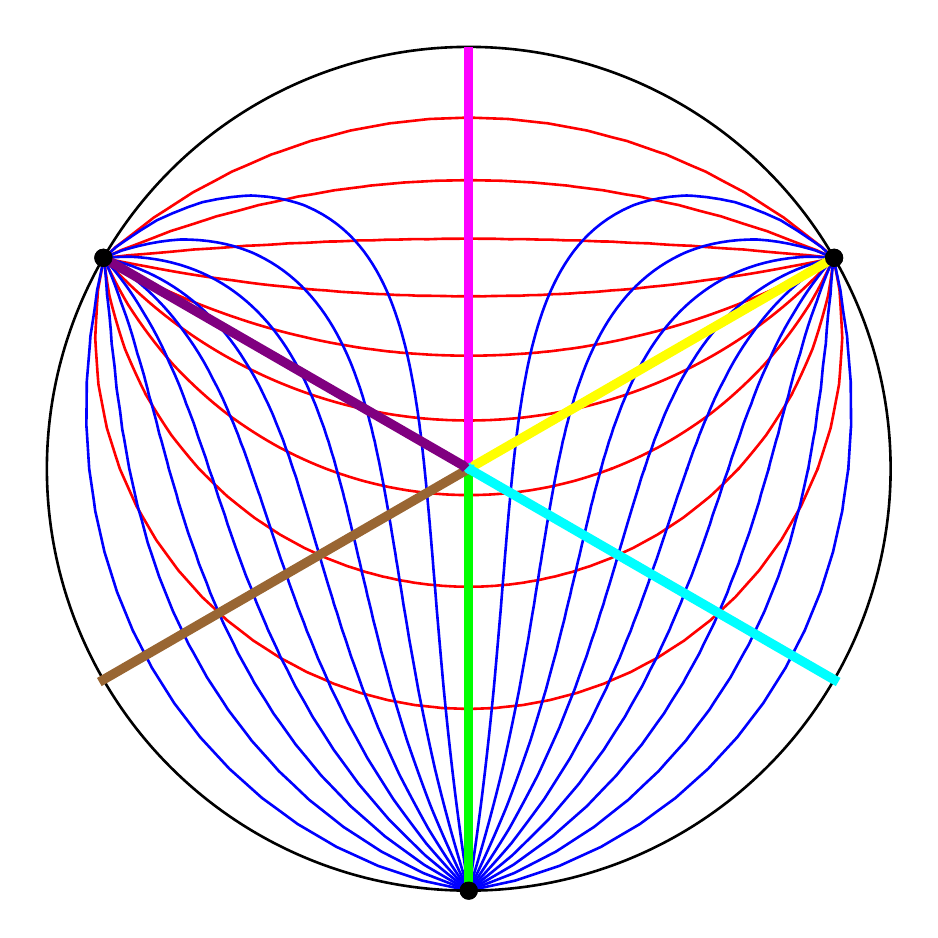}
	}	\hskip 5mm
	\subfigure[][]{\label{fig:YN-charges}
	\begin{tikzpicture}
		\node at (0,0) {\includegraphics[width=0.30\linewidth]{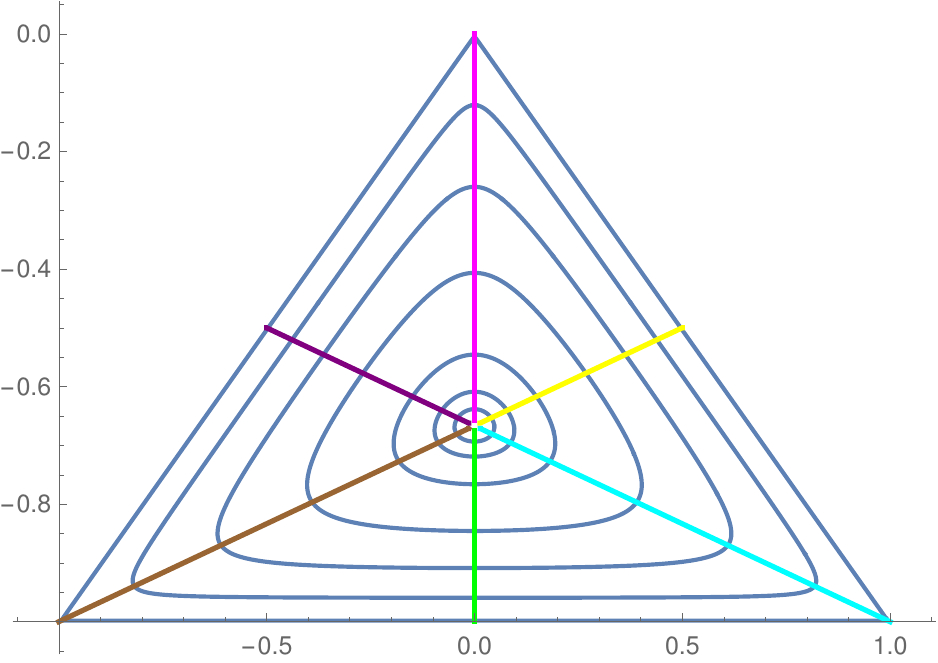}};
		\node[rotate=90] at (-2.6,0) {\scriptsize $N_{\rm F1}/N$};
		\node at (0,-1.9) {\scriptsize $N_{\rm D1}/N$};
	\end{tikzpicture}
	}
	\caption{Left: Brane web for the $Y_N$ theory. Center: Supergravity solution on the disc, with $(N_{\rm D1},N_{\rm F1})$ charges for the $Y_N$ solution and the D3-brane embeddings color-coded as in fig.~\ref{fig:TN}.
	Right: D3-brane embeddings in the space of $(N_{\rm D1},N_{\rm F1})$ charges carved out by curves of constant radius on the disc.
		\label{fig:YN}}
\end{figure}
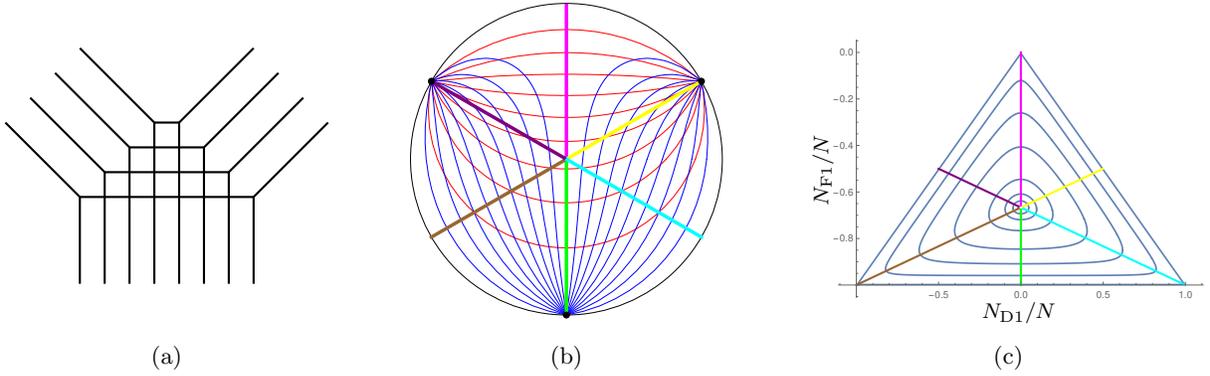

We start with the $T_N$ and $Y_N$ theories. The 5d $T_N$ theories introduced in \cite{Benini:2009gi} are realized by junctions of $N$ D5, $N$ NS5 and $N$ $(1,1)$ 5-branes (fig.~\ref{fig:TN-web}). The $Y_N$ theories of \cite{Bergman:2018hin} correspond to a junction of $2N$ NS5 branes, $N$ $(1,1)$ 5-branes and $N$ $(1,-1)$ 5-branes (fig.~\ref{fig:YN-web}).
The functions $\cA_\pm$ and $\cG$ realizing holographic duals for these theories were given explicitly in \cite{Uhlemann:2020bek}, as
\begin{align}
	 \cA_\pm^{T_N}&=\frac{3N}{8\pi} \left[\pm \ln(w-1)+i\ln(2w)+(\mp 1-i)\ln(w+1)\right]~,
	 \nonumber\\
	  \cA^{Y_N}_\pm&=\frac{3N}{8\pi} \left[(\pm 1+i)\ln(w-1)+(\pm 1-i)\ln(w+1)\mp 2\ln(2w)\right]~,
\label{eq:cA-TN-YN}
\end{align}
where $w$ is a complex coordinate on the upper half plane and $2\pi\alpha^\prime=1$.
The poles on $\partial\Sigma$ are at $w\in\lbrace 0,\pm 1\rbrace$, and encode the charges of the 5-brane junction via (\ref{eq:charges}).
The two supergravity solutions are related by a combination of $SL(2,\RR)$ transformation and rescaling of the charges, as discussed in \cite{Bergman:2018hin}.
The free energies obtained holographically were matched to field theory computations in \cite{Fluder:2018chf,Uhlemann:2019ypp}.

Since the equation governing the D3-brane embedding is invariant under $SL(2,\RR)$ transformations, the solutions for the D3-brane embeddings in $T_N$ and $Y_N$ can be discussed together.
The functions $\cG$ for $T_N$ and $Y_{N/\sqrt{2}}$ are given by
\begin{align}\label{eq:cG-TN}
	\cG_{T_N}=\cG_{Y_{N/\sqrt{2}}}&=\frac{9}{8\pi^2}N^2D\left(\frac{2w}{w+1}\right)~,
\end{align}
where $D$ is the Bloch-Wigner function defined by
\begin{align}\label{eq:D-def}
	D(u)&=\Im\left[\Li_2(u)+\ln(1-u)\ln |u|\right]~.
\end{align}
Since $\cG$ depends on $N$ only through the overall coefficient in (\ref{eq:cG-TN}), the D3-brane BPS equation in (\ref{eq:BPS}) is independent of~$N$.

We start with the embedding describing conformal defects. With constant $w$, (\ref{eq:BPS}) has one solution.
The solution and the on-shell action obtained from (\ref{eq:D3-conf-action}) are given by
\begin{align}\label{eq:TN-conf}
	w_c&=\frac{i}{\sqrt{3}}~, & S_{\rm D3}&=\frac{27}{8\pi^3}N^2\Vol_{AdS_4}\,\Im{\rm Li}_2\big(e^{i\pi/3}\big)~.
\end{align}
One can do a conformal transformation to global $AdS_6$ coordinates, with the D3-brane embedding unchanged, to describe an $S^3$ defect in the 5d SCFT on $S^5$.
The action in (\ref{eq:TN-conf}) with $\Vol_{AdS_4}$ as below (\ref{eq:D3-conf-action}) then gives the contribution of an $S^3$ defect to the free energy on $S^5$ for the $T_N$ and $Y_{N/\sqrt{2}}$ theories.
The orientation of the D3-brane in the (789) direction in the notation of sec.~\ref{sec2}  corresponds to the position of the probe D3-brane on $S^2$. A two-sided defect, as in fig.~\ref{fig:two-side-conf}, can be realized by adding a second probe D3-brane on the antipodal point on $S^2$.

Non-conformal embeddings can be obtained by inverting (\ref{eq:BPS-2}). Using the explicit expression for $\cA_\pm$ for the $T_N$ theory in (\ref{eq:cA-TN-YN}), the real and imaginary parts of (\ref{eq:BPS-2}) lead to
\begin{align}
	\ln\left|\frac{w-1}{w+1}\right|^2&=m_1 e^{-r}~,
	&
	\ln\left|\frac{2w}{w+1}\right|^2&=m_2 e^{-r}~,
\end{align}
where $m_1$ and $m_2$ are, respectively, related to the real and imaginary part of $m$ in (\ref{eq:BPS-2}) by rescaling.
The general solution in $\Sigma$, with Fefferman-Graham radial coordinate $z$,  is given by
\begin{align}\label{eq:TN-sol}
 w&=	\frac{e^{m_1z}-1-i \sqrt{4e^{m_1 z}-\left(e^{m_1z}-e^{m_2z}+1\right)^2}}{e^{m_2 z}-2 e^{m_1 z}-2}~,
 &
 z&=e^{-r}~.
\end{align}
This is a family of curves which connect the conformal point $w_c$ to points on the boundary of $\Sigma$, reaching the boundary when the square root vanishes.

Embeddings with a simple form in the upper half plane, which follow a circle or straight line connecting the point $w_c=i/\sqrt{3}$ to one of the poles, can be found from the fact that $\cG$ is invariant under reflection across the imaginary axis: If the initial departure from the conformal point $w_c$ has vanishing real part this is preserved along the flow,
leading to the solutions (\ref{eq:TN-sol}) with $m_1=0$, 
\begin{align}
	w_{1}(z)&=\frac{i}{\sqrt{4e^{-m_2 z}-1}}~.
\end{align}
The supergravity solution further has a $\ZZ_3$ group of $SL(2,\RR)$ transformations of the upper half plane which can be combined with Type IIB $SL(2,\RR)$ transformations to form symmetries of the solution. 
Since the D3-brane BPS condition is invariant under Type IIB $SL(2,\RR)$ transformations, this leads to two related branches of solutions connecting $w_c$ to the other two poles,
\begin{align}
	w_2(z)&=\frac{1+w_1(z)}{1-3w_1(z)}~,
	&
	w_3(z)&=\frac{1+w_2(z)}{1-3w_2(z)}~.
\end{align}
These D3-brane embeddings resemble the string embeddings discussed in \cite{Bergman:2018hin}. 
The form of the embeddings is not symmetric in $m_2\rightarrow -m_2$: 
For $m_2<0$ the D3-brane reaches all the way into the IR region of $AdS_6$, where $z\rightarrow \infty$, without reaching the pole on the boundary of $\Sigma$. 
For $m_2>0$, on the other hand, the D3-brane reaches a regular point on the boundary of $\Sigma$ at $m_2z=\ln 4$. 

As the D3-brane reaches the boundary of $\Sigma$, it can not just end at the corresponding value of the $AdS_6$ radial coordinate~$z$:
since the D3-brane does not wrap any part of the internal space, one can not obtain a smooth worldvolume without boundary by shrinking an internal cycle.
However, since the $S^2$ in which the D3-brane is localized at a point collapses on the boundary of $\Sigma$, 
one can connect the embedding to a second D3-brane, described by the same profile $w(r)$ but located at the antipodal point of $S^2$.
Together they form a smooth D3-brane worldvolume, with the branches at the two antipodal points of $S^2$ corresponding to the two branches of the two-sided defects in fig.~\ref{fig:defect-D3}.
In the UV region of $AdS_6$, where $z$ is small, the two D3-branes are close to the conformal embedding on $\Sigma$.
As one moves into the IR region of $AdS_6$ both branches approach the boundary of $\Sigma$, where they meet for $zm_2=\ln 4$. At that point they disappear from the $AdS_6$ perspective.

In summary, mass deformations with $m_2<0$ can be realized for one-sided defects in the sense of fig.~\ref{fig:defect-D3}.
They can also be realized for two-sided defects, which simply corresponds to adding a second embedding at the antipodal point of $S^2$.
Mass deformations with $m_2>0$ can only be realized for two-sided defects in the sense of fig.~\ref{fig:defect-D3} with two branches of D3-brane embeddings, and the mass deformations completely gap the defect in the IR. This realizes brane configurations of the type shown in fig.~\ref{fig:two-side-non-conf}.

\begin{figure}
	\includegraphics[width=0.28\linewidth]{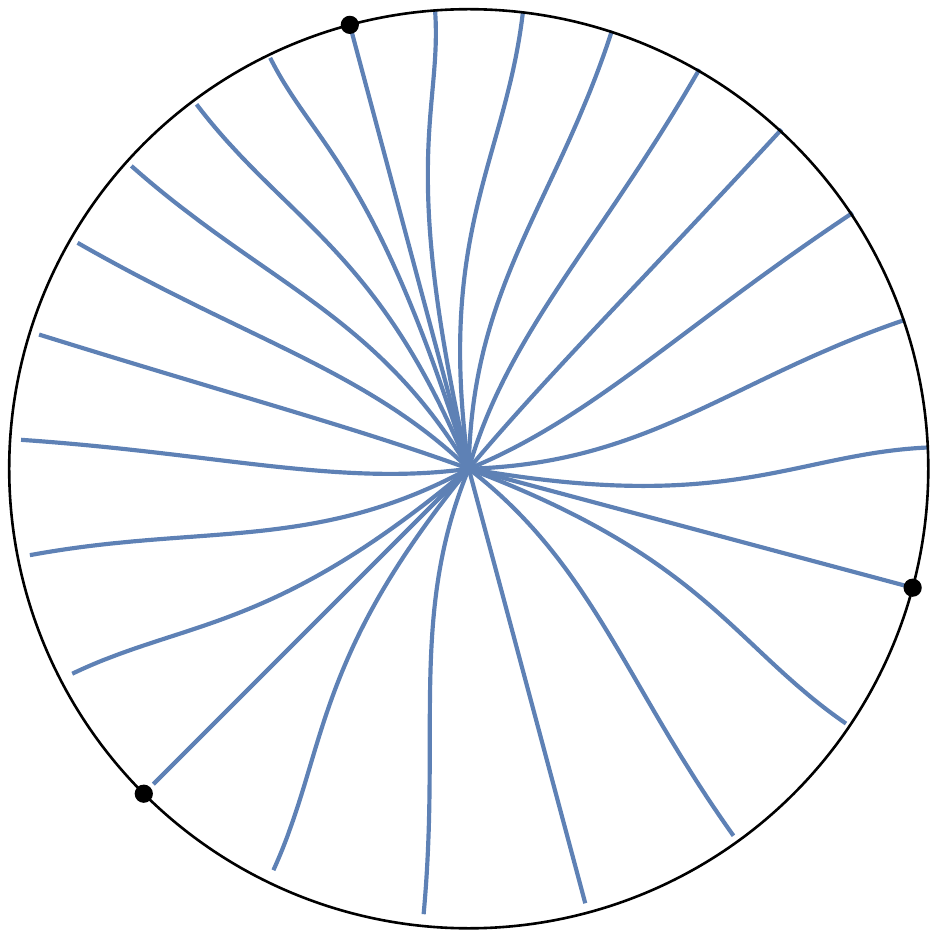}
	\hskip 20mm
	\begin{tikzpicture}
		\node at (0,0) {\includegraphics[width=0.26\linewidth]{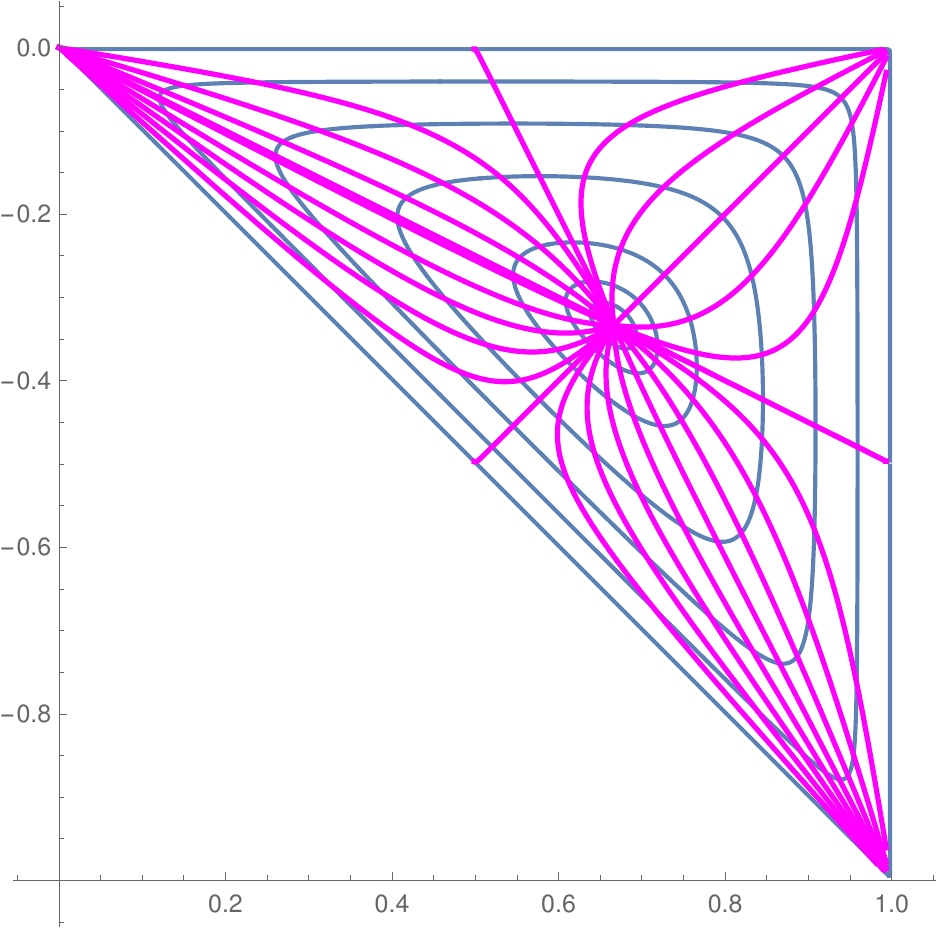}};
		\node[rotate=90] at (-2.3,0) {\scriptsize $N_{\rm F1}/N$};
		\node at (0,-2.25) {\scriptsize $N_{\rm D1}/N$};
	\end{tikzpicture}
	\caption{General D3-brane embeddings for $T_N$. Generic embeddings reach the boundary of $\Sigma$ at regular points. Only the fine-tuned ones approach the poles. On the right in the charge lattice, where the edges represent the poles. Generic embeddings approach the corners, the fine-tunes ones approach the edges.\label{fig:TN-gen}}
\end{figure}

The embeddings are illustrated in figs.~\ref{fig:TN-disc} and \ref{fig:TN-charges} for the $T_N$ solution. Fig.~\ref{fig:TN-disc} illustrates the supergravity solution with $\Sigma$ mapped to the unit disc with coordinate $u$ via
\begin{align}\label{eq:u-def}
	\frac{\sqrt{3}w-i}{\sqrt{3}w+i}&=e^{-i\pi/4}u~.
\end{align}
This maps the $Z_3$-symmetric point $w=i/\sqrt{3}$ to the center of the disc.
The poles of $\partial\cA_\pm$, where the 5-brane stacks are located, are indicated as solid circles on the boundary,
and the phase of $u$ in (\ref{eq:u-def}) is chosen such that the locations of the poles roughly line up with the external 5-brane stacks in the brane web in fig.~\ref{fig:TN-web}.\footnote{For 3-pole solutions one could choose the mapping such that the poles precisely reflect the angles of the external 5-branes in the brane web, as done in \cite{Uhlemann:2020bek}. This is not possible in general.}
As shown in \cite{Uhlemann:2020bek} by considering Wilson loops represented by different D3-brane embeddings, each point of $\Sigma$ can be identified with a face of the associated 5-brane web. Namely, each point on $\Sigma$ can be assigned coordinates $(N_{\rm D1},N_{\rm F_1})$  given by
\begin{align}
	N_{\rm F1}+i N_{\rm D1}&=\frac{4}{3}\left(\cA_+ +\bar \cA_-\right)~.
\end{align}
These coordinates translate to coordinates on the brane web, with $N_{\rm D1}$ labeling the faces in the horizontal direction and $N_{\rm F1}$ labeling the faces in the vertical direction. The space of charges carved out when moving along $\Sigma$ is shown by the blue closed curves in fig.~\ref{fig:TN-charges}, which correspond to curves of constant radius on the disc. The points in the triangle correspond to the closed faces of the brane web in fig.~\ref{fig:TN-web}, and the shape represents the associated Newton polygon.

In fig.~\ref{fig:TN-disc} curves of constant $N_{\rm D1}$/$N_{\rm F1}$ are shown in red/blue. The defect embeddings are the straight lines connecting the center of the disc to boundary points, with each branch shown in a different color. The lines ending at the poles describe non-conformal D3-brane defects that reach into the IR region of $AdS_6$ -- these embeddings never actually reach the poles. The lines ending at regular boundary points represent a pair of embeddings at antipodal points of the $S^2$ which are joined at the boundary of $\Sigma$, and describe two-sided defects.  These embeddings describe defects that are gapped in the IR.
Fig.~\ref{fig:TN-charges} shows how the D3-brane embeddings $w_1, w_2, w_3$ trace through the $(N_{\rm D1},N_{\rm F1})$ coordinates. The quiver gauge theory deformation of the $T_N$ theory is given by
\begin{align}
	[2]-SU(2)-SU(3)-\ldots SU(N-1)-[N]~.
\end{align}
The discrete symmetries of the supergravity solution and brane web do not translate to simple symmetries of the quiver gauge theory.
This will be different for the $Y_N$ theory.

More general embeddings (\ref{eq:TN-sol}) are shown in fig.~\ref{fig:TN-gen}.
The general picture is that, for generic choices of the phase of $m_1+im_2$, the D3-brane embedding reaches the boundary of $\Sigma$ at a regular point, so that it has to be combined with a second branch to form a two-sided defect. 
There are three distinguished choices of the phase for which the defect approaches the boundary of $\Sigma$ at a pole, so that a one-sided embedding can be realized and extends all the way into the IR.
This mirrors the discussion of sec.~\ref{sec2}, where the one-sided embeddings can only be moved along specific directions in the (56) plane in which the external 5-branes of the junction extend, while two-sided D3-brane defects can be moved arbitrarily into the (56) plane.

For the $Y_N$ solution the D3-brane embeddings take the same form in $\Sigma$, though the brane web and gauge theories are different. 
The brane web and supergravity solution are shown in fig.~\ref{fig:YN}. 
The $Y_N$ theories have two S-dual gauge theory deformations that differ in their form. 
The one corresponding to the brane web in fig.~\ref{fig:YN-web} is given by
\begin{align}\label{eq:YN-quiver-2}
	[2]-SU&(2)-\ldots-SU(N\,{-}\,1)-SU(N)^{}_{\pm 1}-SU(N\,{-}\,1)-\ldots-SU(2)-[2]~.
\end{align}
Along the first/second ellipsis the rank of the gauge groups increases/decreases in steps of one, and the central node has a Chern-Simons term.
The quiver is symmetric under reflection across the central node, and this symmetry is realized  in the supergravity solution as reflection across the vertical diameter of the disc in fig.~\ref{fig:YN-disc}.
The form of the D3-brane embeddings suggests that the mass deformation described by $w_1$ is symmetric under reflection of the quiver across the central node. 
The Chern-Simons term explains the lack of symmetry under $m_2\rightarrow -m_2$.

\subsection{\texorpdfstring{$+_{N,M}$}{+[N,M]} and \texorpdfstring{$X_{N,M}$}{X[N,M]} theories}

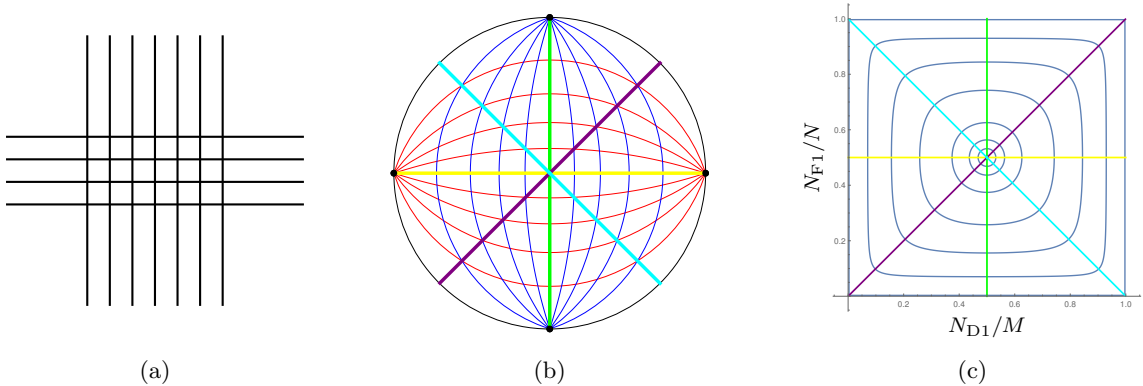
\begin{figure}
	\subfigure[][]{\label{fig:plus-web}
		\begin{tikzpicture}[scale=0.6]
			\foreach \i in {-1.5,-1.0,-0.5,0,0.5,1,1.5}{
				\draw[thick] (\i,-3) -- (\i,3);
			}
			\foreach \j in {-0.75,-0.25,0.25,0.75}{
				\draw[thick] (3.3,\j) -- (-3.3,\j);
			}
			\node at (0,-3.7) {};
			
		\end{tikzpicture}
	}	\hskip 6mm
	\subfigure[][]{\label{fig:plus-disc}
		\includegraphics[width=0.28\linewidth]{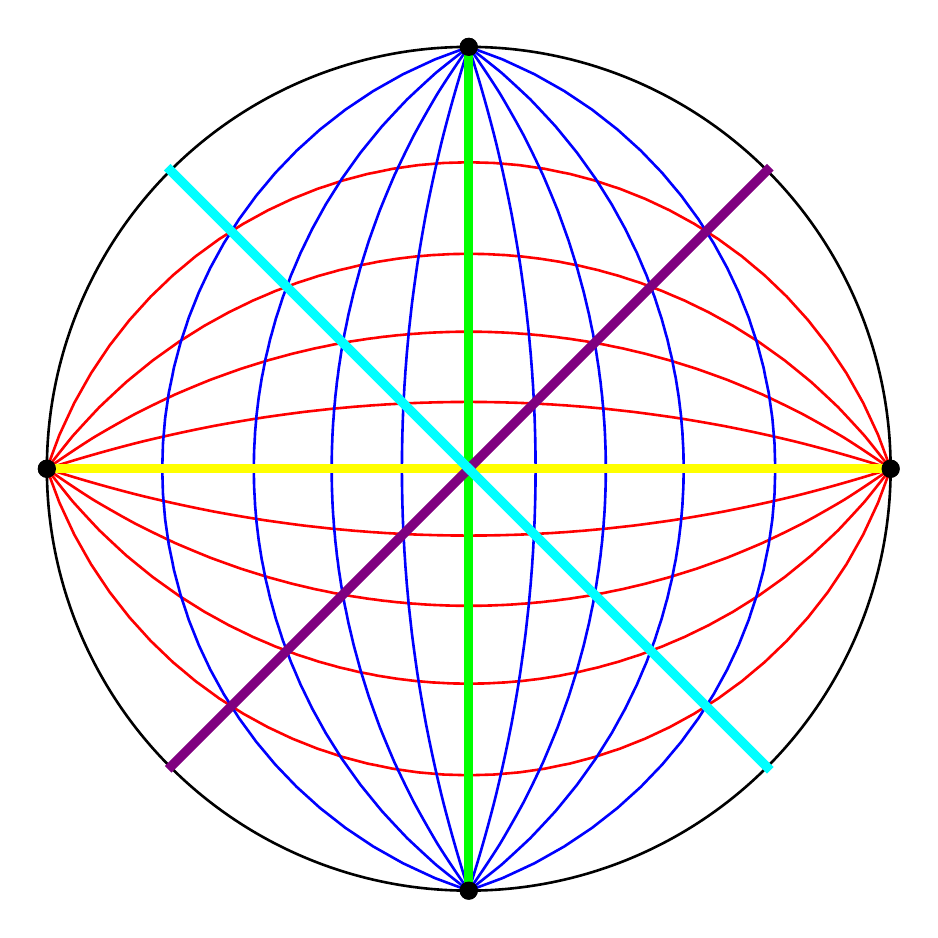}
	}	\hskip 6mm
	\subfigure[][]{\label{fig:plus-charges}
		\begin{tikzpicture}
			\node at (0,0) {\includegraphics[width=0.25\linewidth]{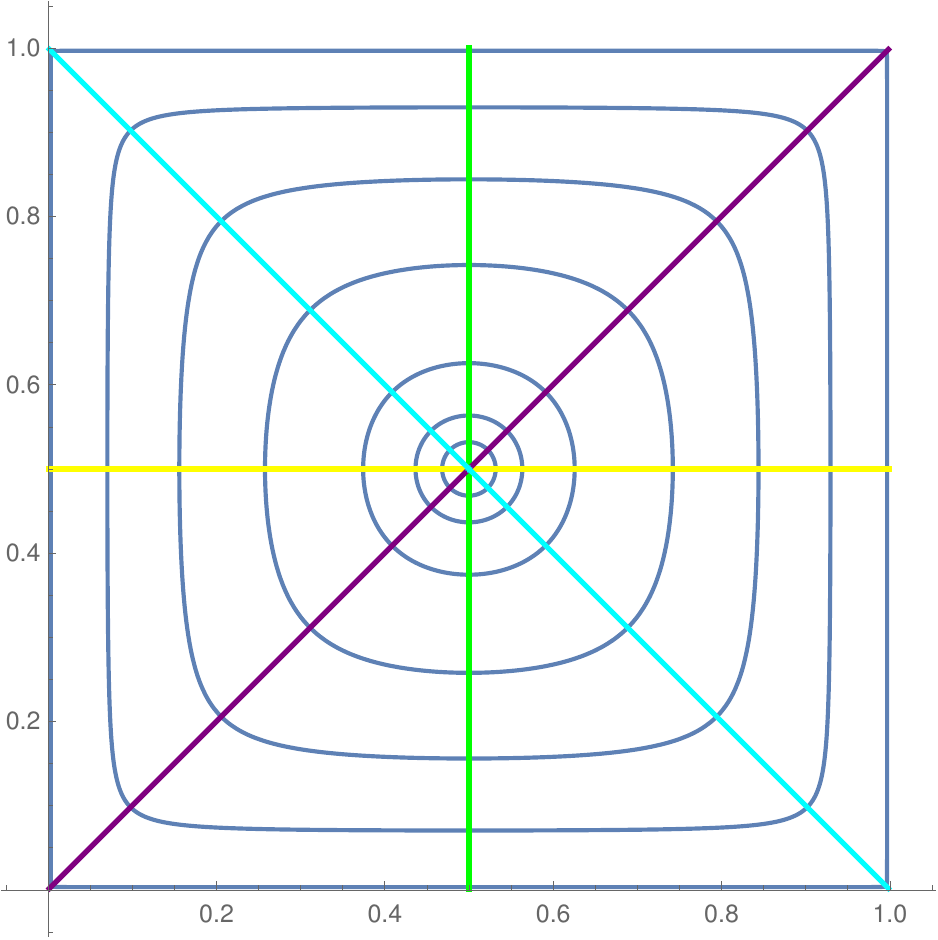}};
			\node[rotate=90] at (-2.3,0) {\scriptsize $N_{\rm F1}/N$};
			\node at (0,-2.25) {\scriptsize $N_{\rm D1}/M$};
		\end{tikzpicture}
		
	}
	\caption{Left: brane web for the $+_{N,M}$ theory. Center: supergravity solution on the disc, with curves of constant $N_{\rm F1}$, $N_{\rm D1}$ in red, blue. The D3-brane embedding $w_1$ is shown in yellow, $w_2$ in green, $w_3$ in purple, $w_4$ in cyan.
		Right: $(N_{\rm D1},N_{\rm F1})$ along curves of constant radius on the disc and the D3-brane embeddings.\label{fig:plus}}
\end{figure}

The $+_{N,M}$ theories are realized by intersections of $N$ D5 and $M$ NS5-branes (fig.~\ref{fig:plus-web}), and were studied already in \cite{Aharony:1997bh}.
The $X_{N,M}$ theories of \cite{Bergman:2018hin} correspond to intersections of $N$ $(1,-1)$ 5-branes and $M$ $(1,1)$ 5-branes. Similar to the relation between the $T_N$ and $Y_N$ theories, the supergravity solutions are related by a combination of Type IIB $SL(2,\RR)$ transformations and charge rescaling.
The functions $\cA_\pm$ are
\begin{align}
	\cA_\pm^{+_{N,M}}&=\frac{3}{8\pi}\left[iN\left(\ln(2w-1)-\ln(w-1)\right)\pm M \left(\ln (3w-2)-\ln w\right)\right]~,
	\nonumber\\
	 \cA_\pm^{X_{N,M}}&=\frac{3}{8\pi}\left[
	(\pm 1+i)M\left(\ln(3w-2)-\ln w\right)+(\pm 1-i)N\left(\ln(w-1)-\ln(2w-1)\right)
	\right]\,,
\end{align}
where $w$ is a complex coordinate on the upper half plane.
The poles where the external 5-branes emerge are at $w\in\lbrace 0,\tfrac{1}{2},\tfrac{2}{3},1\rbrace$.
Due to the $SL(2,\RR)$ relation between the solutions, the functions $\cG$ are closely related. For $+_{N,M}$ and $X_{N/\sqrt{2},M/\sqrt{2}}$ \cite{Uhlemann:2020bek}
\begin{align}\label{eq:cG-plus}
	\cG_{+_{N,M}}=\cG_{X_{N/\sqrt{2},M/\sqrt{2}}}&=\frac{9}{8\pi^2}NM\left[D\left(\frac{3w-2}{w}\right)+D\left(\frac{w}{2-3w}\right)\right]~.
\end{align}
The free energy obtained from the supergravity solution was matched to field theory computations in \cite{Fluder:2018chf,Uhlemann:2019ypp} for the $+_{N,M}$ theory and in \cite{Uhlemann:2020bek} for $X_{N,N}$.

Since $N$ and $M$ only appear as combined overall factor in (\ref{eq:cG-plus}), the D3-brane BPS equation (\ref{eq:BPS}) is independent of $N$ and $M$.
The embedding describing a conformal defect with constant $w(r)$ can again be obtained from (\ref{eq:BPS-massless}). 
The solution $w_c$ and the on-shell action for $+_{N,M}$ and $X_{N/\sqrt{2},M/\sqrt{2}}$ are given by
\begin{align}
	w_c&=\frac{3+i}{5}~, & S_{\rm D3}&=\frac{27C}{4\pi^3}MN \Vol_{\rm AdS_4}~,
\end{align}
where $C\approx 0.916$ is Catalan's constant.
With the renormalized volume of $AdS_4$ as given below (\ref{eq:D3-conf-action}) this yields the contribution of an $S^3$ defect to the free energy of the SCFTs on $S^5$.

Non-conformal defects are obtained by inverting (\ref{eq:BPS-2}). With the expressions for $\cA_\pm$ for the $+_{N,M}$ theory, the real and imaginary parts of (\ref{eq:BPS-2}) lead to
\begin{align}
	\ln\left|\frac{3w-2}{w}\right|&=m_1e^{-r}~, & \ln \left|\frac{2w-1}{w-1}\right|&=m_2 e^{-r}~.
\end{align}
Solving for $w(r)$ leads to the general form of the embeddings
\begin{align}\label{eq:plus-sol-w}
w&=\frac{2}{3-i e^{m_1z-i \sin ^{-1}(\cosh (m_1z) \tanh (m_2 z))}}~, & z&=e^{-r}~.
\end{align}
These are again curves connecting the conformal point $w_c$ to points on the boundary of $\Sigma$.
The form of the embeddings is symmetric in $m_1\rightarrow -m_1$ and $m_2\rightarrow -m_2$.

The solutions again simplify for flows along curves that are invariant under discrete symmetries of the background solution.
Concretely, one can use an $SL(2,\RR)$ transformation on the upper half plane to map the poles to $\lbrace -1,0,1,\infty\rbrace$ and then find embeddings along the imaginary axis. This corresponds to $m_2=0$ in (\ref{eq:plus-sol-w}) and leads to
\begin{align}
	w_1&=\frac{2}{3-i e^{m_1  z}}~.
\end{align}
These embeddings connect the point $w_c$ to the two D5-brane poles. 
In the IR the embeddings $w_1$ approach the D5-brane poles on $\Sigma$ but never reach them, so that the D3-branes extend all the way into the IR region of $AdS_6$.
In fig.~\ref{fig:plus-disc} these embeddings correspond to the horizontal yellow line.
They describe one-sided defects in the sense of sec.~\ref{sec2} with the D3-branes displaced from the junction point along the D5-branes.
A second family of solutions can be obtained from $w_1$ by an $SL(2,\RR)$ transformation of the upper half plane which cyclically permutes the positions of the poles. This leads to the solutions (\ref{eq:plus-sol-w}) with $m_1=0$,
\begin{align}
	w_2&=1-\frac{1}{2+i e^{m_2 z}}~.
\end{align}
These embeddings correspond to mass deformations of the conformal defect that are S-dual to the mass deformations leading to the embeddings $w_1$; they connect the point $w_c$ to the NS5-brane poles.
In fig.~\ref{fig:plus-disc} they correspond to the green horizontal line.

\begin{figure}
	\includegraphics[width=0.28\linewidth]{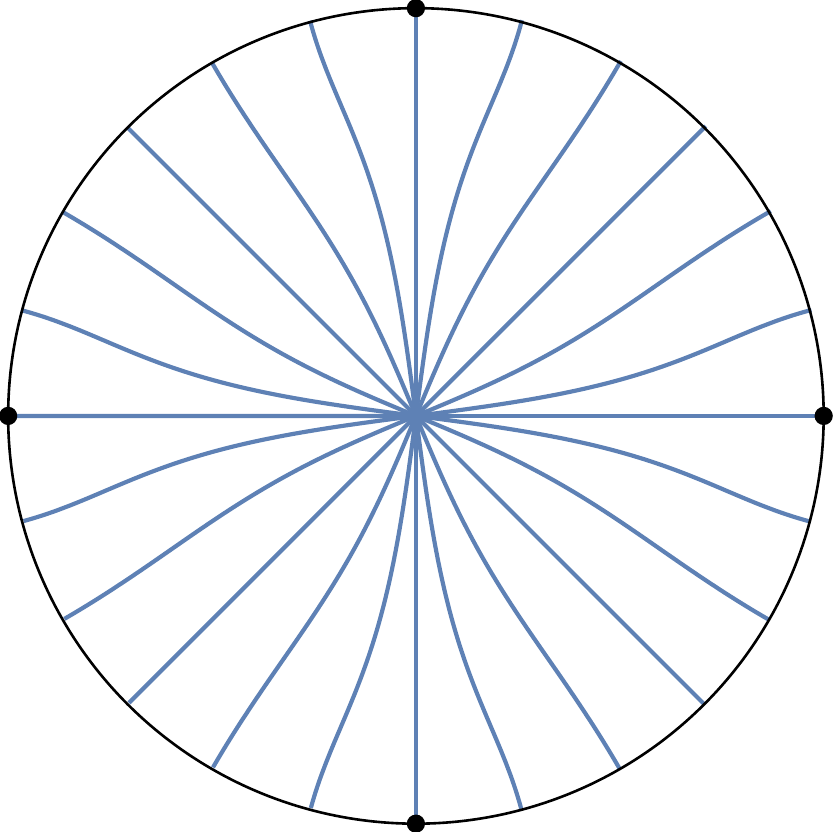}
	\hskip 20mm
	\begin{tikzpicture}
		\node at (0,0) {\includegraphics[width=0.26\linewidth]{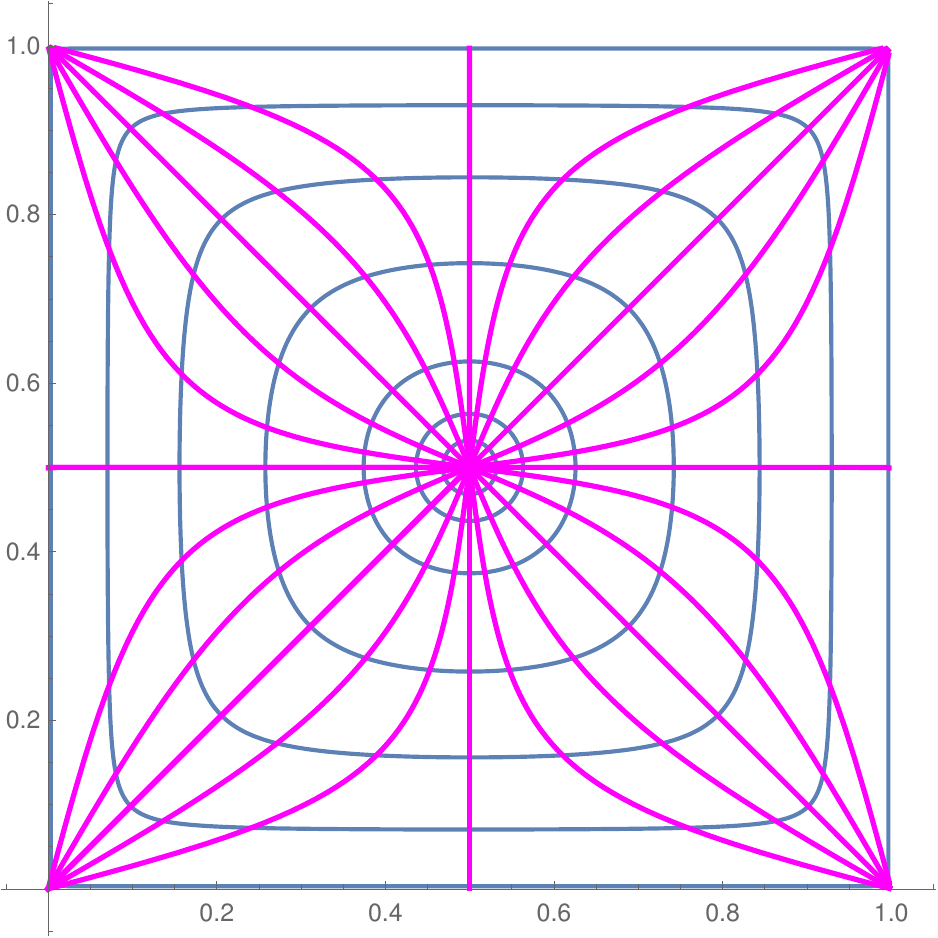}};
		\node[rotate=90] at (-2.3,0) {\scriptsize $N_{\rm F1}/N$};
		\node at (0,-2.25) {\scriptsize $N_{\rm D1}/M$};
	\end{tikzpicture}
	\caption{More general D3-brane embeddings for $+_{N,M}$. Generic embeddings reach the boundary of $\Sigma$ at regular points, only $w_{1/2}$ approach the poles. On the right in the $(N_{\rm D1},N_{\rm F1})$ charges, where the edges represent the poles. Generic embeddings approach the corners, the embeddings $w_{1/2}$ approach the edges.\label{fig:plus-gen}}
\end{figure}

Two more simple solutions correspond to the diagonal lines in fig.~\ref{fig:plus-disc}.
They can be obtained by performing $SL(2,\RR)$ transformations of the upper half plane that map the poles to locations that are  symmetric with respect to reflection across the imaginary axis, and then making an ansatz for imaginary embeddings.
This corresponds to $m_1=\pm m_2$ in (\ref{eq:plus-sol-w}), leading to
\begin{align}
	w_{3/4}&=\frac{4}{6\pm 1\mp e^{2m_1z}-i \sqrt{6 e^{2m_1 z}-e^{4 m_1 z}-1}}~.
\end{align}
In fig.~\ref{fig:plus-disc} the purple line shows $w_3$ and the cyan line $w_4$.
The solutions $w_3$ and $w_4$ reach the boundary at $2|m_1|z=\ln(3+2\sqrt{2})$.
They do not extend all the way into the IR region of $AdS_6$ and have to be combined with a second branch of the same embedding on the antipodal point on $S^2$, to form a two-sided defect in the sense of sec.~\ref{sec2}.

The general picture is similar to the one for the $T_N$ and $Y_N$ theories, and illustrated in fig.~\ref{fig:plus-gen}. For generic choices of $m_1+im_2$, the embedding reaches the boundary of $\Sigma$ at finite $AdS_6$ radial coordinate, and has to be combined with a second branch on the antipodal point of $S^2$ to form a regular embedding. The defect degrees of freedom are gapped in the IR. 
For discrete choices of the phase of $m_1+im_2$, in this case four, the non-conformal embeddings approach the poles without ever reaching them, and can describe one-sided defects. The embedding then reaches all the way into the IR of $AdS_6$ and the defect contains light degrees of freedom.

The quiver gauge theory deformations of the $+_{N,M}$ junction reflect the $\ZZ_2$ symmetries in the supergravity solution.
The quiver is given by 
\begin{align}\label{eq:plus-quiver}
	[N]-SU(N)-\ldots -SU(N)-[N]~,
\end{align}
with a total of $M-1$ gauge nodes. The S-dual quiver deformation, corresponding to a ninety degree rotation of the brane web, has the same form but with $N$ and $M$ exchanged.
The plots in fig.~\ref{fig:plus-charges} suggest that the defects described by the embedding $w_2$ correspond to mass deformations that are symmetric under reflection of the quiver (\ref{eq:plus-quiver}) across the central node.
Analogous comments hold for the defects described by the embedding $w_1$ in the S-dual quiver.
The analog of fig.~\ref{fig:plus-charges} for the $X_{N,M}$ theory follows from a 45 degree rotation, similarly to the relation between the $T_N$ and $Y_N$ theories before (the horizontal and vertical axes in fig.~\ref{fig:plus-charges} are scaled differently for $N\neq M$, and the rotation is to be performed with unrescaled axes).

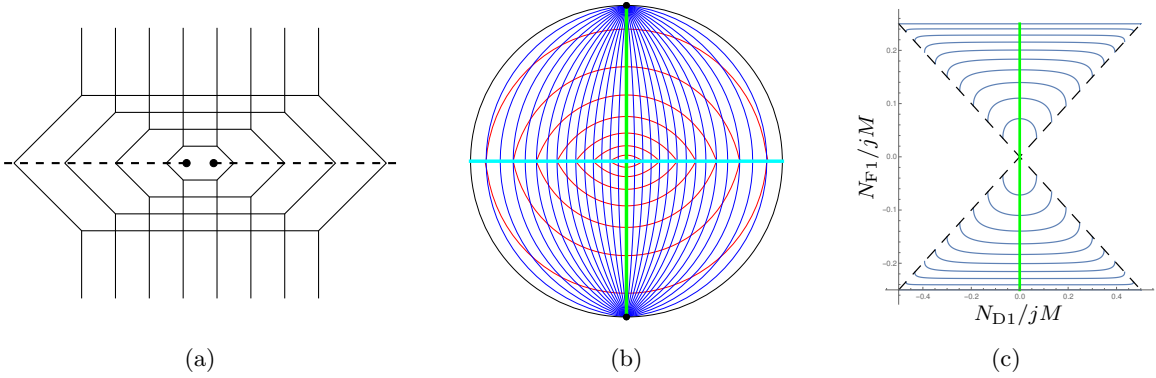
\begin{figure}
	\subfigure[][]{\label{fig:I-web}
		\begin{tikzpicture}[scale=0.9]
  \draw[dashed,black,thick] (0.2,0) -- (3,0);
\draw[dashed,black,thick] (-0.2,0) -- (-3,0);
\draw[fill=black] (0.2,0) circle (1.5pt);
\draw[fill=black] (-0.2,0) circle (1.5pt);

\draw (-0.5,0) -- (-0.25,0.25) -- (0.25,0.25) -- (0.5,0) -- (0.25,-0.25) -- (-0.25,-0.25) --(-0.5,0);
\draw (-0.25,0.25) -- (-0.25,2);
\draw (0.25,0.25) -- (0.25,2);
\draw (-0.25,-0.25) -- (-0.25,-2);
\draw (0.25,-0.25) -- (0.25,-2);

\draw (-2,0) --(-1.25,0.75) -- (1.25,0.75) -- (2,0) -- (1.25,-0.75) -- (-1.25,-0.75) -- (-2,0);
\draw (-1.25,0.75) -- (-1.25,2);
\draw (1.25,0.75) -- (1.25,2);
\draw (-1.25,-0.75) -- (-1.25,-2);
\draw (1.25,-0.75) -- (1.25,-2);

\draw (-1.25,0) -- (-0.75,0.5) -- (0.75,0.5) -- (1.25,0) -- (0.75,-0.5) -- (-0.75,-0.5) -- (-1.25,0);
\draw (-0.75,0.5) -- (-0.75,2);
\draw (0.75,0.5) -- (0.75,2);
\draw (-0.75,-0.5) -- (-0.75,-2);
\draw (0.75,-0.5) -- (0.75,-2);

\draw (-2.75,0) -- (-1.75,1) -- (1.75,1) -- (2.75,0) -- (1.75,-1) -- (-1.75,-1) -- (-2.75,0);
\draw (-1.75,1) -- (-1.75,2);
\draw (1.75,1) -- (1.75,2);
\draw (-1.75,-1) -- (-1.75,-2);
\draw (1.75,-1) -- (1.75,-2);

\node at (0,-2.4) {};
		\end{tikzpicture}
	}	\hskip 3mm
	\subfigure[][]{\label{fig:I-disc}
		\includegraphics[width=0.28\linewidth]{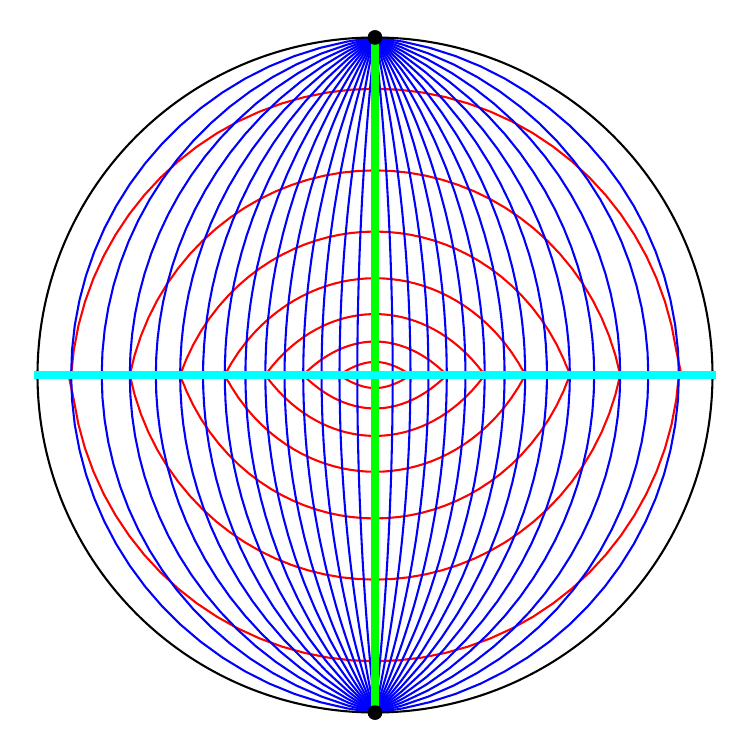}
	}	\hskip 3mm
	\subfigure[][]{\label{fig:I-charges}
		\begin{tikzpicture}
			\node at (0,0) {\includegraphics[width=0.22\linewidth]{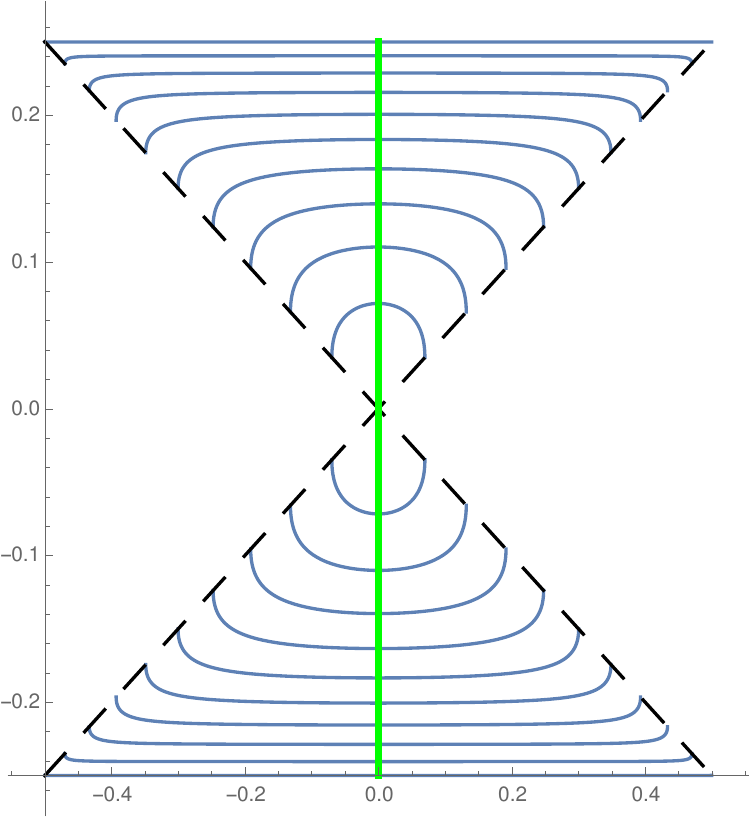}};
			\node[rotate=90] at (-2.0,0) {\scriptsize $N_{\rm F1}/jM$};
			\node at (0,-2.1) {\scriptsize $N_{\rm D1}/jM$};
		\end{tikzpicture}
	}
	\caption{Left: brane web for the $I_{M,j}$ theories, with $j=2$. Center: supergravity solution on the disc, with $v_1$ in green, $v_2$ in cyan, and at the center the puncture with two branch cuts along the horizontal diameter. 
	In the $(N_{\rm D1},N_{\rm F1})$ charges on the right the $v_2$ embedding is on the branch cuts, shown as dashed lines.
		\label{fig:I}}
\end{figure}

\subsection{\texorpdfstring{$\mathsf{I}_{M,j}$}{I[M,j]} theories}

As a last example we consider a 5-brane junction with 7-branes.
It is composed of two stacks of $M$ NS5-branes on a system of $2j$ D7-branes.
The D7-branes are split into two groups of $j$ D7-branes which have their branch cuts oriented in opposite directions (fig.~\ref{fig:I-web}).
This is a special case of the $+_{N,M,j,k}$ theories discussed in \cite{Chaney:2018gjc}. 
Orientifolds, dubbed $\mathsf{I}_{M,j}^\pm$, were discussed in \cite{Uhlemann:2019lge}.
The functions $\cA_\pm$ can be written as (see (3.4) of \cite{Uhlemann:2019lge})
\begin{align}\label{eq:cA-I}
	\cA_\pm&=\frac{3M}{8\pi}\left[
	\pm \ln\left(\frac{1+i v}{1-i v}\right)+\frac{j}{\pi}\left(\Li_2(i v)-\Li_2(-i v)-\tanh^{-1}(iv)\ln(-v^2)\right)
	\right],
\end{align}
where $v$ is a complex coordinate on the unit disc. The two poles on the boundary of $\Sigma$ corresponding to the external NS5-brane stacks are at $v=\pm i$, and there are $2j$ D7-branes at the center of the disc.
Half of the D7-branes have their branch cut along the positive real axis, the other half has their branch cut along the negative real axis. The solutions have two $\ZZ_2$ symmetries, corresponding to reflection across the real and imaginary axes.
The quiver gauge theory is given by (4.17) of \cite{Chaney:2018gjc} with $k=j$ and $2N=Mj$.
The free energy was matched to a field theory computation in \cite{Uhlemann:2019lge}.

The quantities $\partial\cG=(\bar \cA_+-\cA_-)\partial\cA_+ +(\cA_+-\bar \cA_-)\partial\cA_-$ and $\kappa^2$ feeding into the BPS equation for the D3-branes are single-valued and both depend on $M$ and $j$ only through an overall coefficient $jM^2$, so that the D3-brane embeddings are independent of $M$ and $j$.
The BPS equation for conformal defects (\ref{eq:BPS-massless}) has one solution. 
The solution and the on-shell action for the D3-brane evaluated on the solution are given by
\begin{align}
	v_c&=0~, &
	S_{\rm D3}&=\frac{189}{32\pi^4}\zeta(3) \Vol_{AdS_4} jM^2~.
\end{align}
The conformal defect is thus localized at the center of the disc, on top of the D7-brane punctures.

Non-conformal embeddings can be obtained from (\ref{eq:BPS-2}). 
With $\cA_\pm$ in (\ref{eq:cA-I}) the real and imaginary parts of (\ref{eq:BPS-2}) lead to
\begin{align}\label{eq:I-BPS-imp}
	\ln\left|\frac{v+i}{v-i}\right|&=m_1 e^{-r}~, &
	\Im\left[\Li_2(i v)-\Li_2(-i v)-i\ln(-v^2)\tan^{-1}(v)\right]&=m_2 e^{-r}~.
\end{align}
The form of the embeddings is symmetric in $m_1\rightarrow -m_1$ and $m_2\rightarrow -m_2$.

The background solution has two $\ZZ_2$ symmetries corresponding to reflection across the real and imaginary axes, so there are two distinguished families of embeddings along these fixed lines.
For embeddings along the imaginary axis the BPS equations simplify since the dilogarithm terms drop out.
They correspond to $m_2=0$ and can be given explicitly as
\begin{align}
	v_1&=i\tanh\left(\frac{m_1z}{4}\right)~, & z&=e^{-r}~.
\end{align}
The embeddings connect the center of the disc to the NS5-brane poles.
They describe one-sided defects and extend all the way into the IR region of $AdS_6$. 
For embeddings along the real axis, with $v_2(r)=\bar v_2(r)$, the first equation in (\ref{eq:I-BPS-imp}) is trivial, and the second equation simplifies to
\begin{align}
	2D(i v_2)&=m_2 e^{-r}~,
\end{align}
where $D$ is the Bloch-Wigner function defined in (\ref{eq:D-def}).
Since $D_2(\pm i)=\pm C$, where $C$ is Catalan's constant, the D3-brane reaches the boundary of $\Sigma$, with $v_2$ reaching $\pm 1$, when $|m_2|e^{-r}=2C$.
These embeddings have to be combined with a second branch on the antipodal point of $S^2$ to form regular embeddings describing two-sided defects. 

\begin{figure}
	\includegraphics[width=0.28\linewidth]{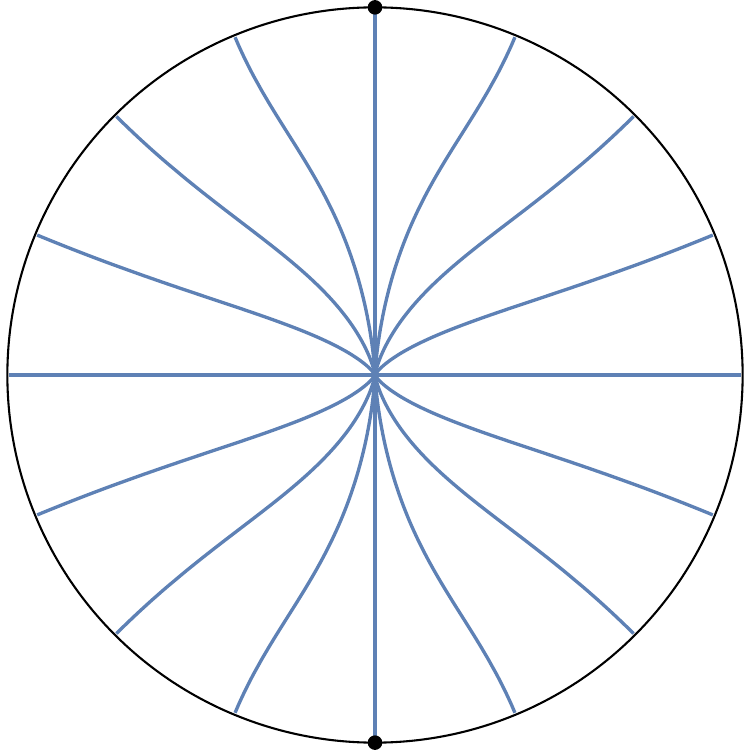}
	\hskip 20mm
	\begin{tikzpicture}
		\node at (0,0) {\includegraphics[width=0.25\linewidth]{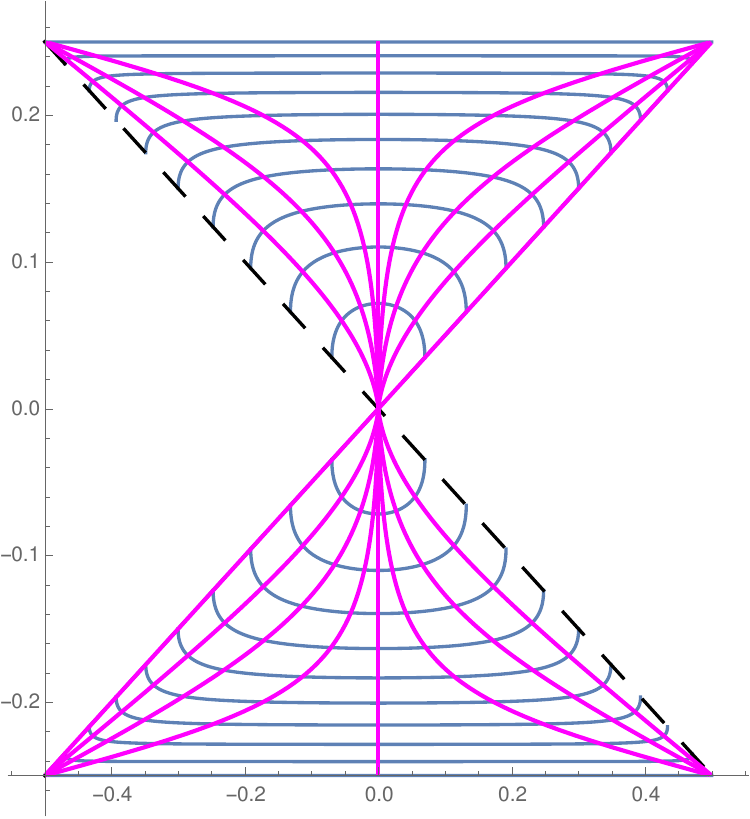}};
		\node[rotate=90] at (-2.3,0) {\scriptsize $N_{\rm F1}/jM$};
		\node at (0,-2.25) {\scriptsize $N_{\rm D1}/jM$};
	\end{tikzpicture}
	\caption{General D3-brane embeddings for $I_{M,j}$. Generic embeddings reach the boundary of $\Sigma$ at regular points, only $v_1$ and $v_2$ approach the poles. On the right in the space of $(N_{\rm D1},N_{\rm F1})$ charges.\label{fig:I-gen}}
\end{figure}

More general solutions are shown in fig.~\ref{fig:I-gen}. As before, generic embeddings reach regular points on the boundary of $\Sigma$ at finite radial coordinate in $AdS_6$, and have to be combined with a second branch on the antipodal point of $S^2$ to form a smooth worldvolume. Only the embeddings $v_1$ and $v_2$ approach the poles and describe one-sided defects.

\section{Discussion}\label{sec4}

We have investigated planar three-dimensional surface defects in 5d SCFTs engineered by the UV limit of $(p,q)$ 5-brane webs. 
The surface defects are realized either by D3-branes ending on the 5-brane junction arising as UV limit of the brane web, or by D3-branes intersecting the plane in which $(p,q)$ 5-branes are at an angle dictated by their charges.
Conformal defects are realized by D3-branes at the point where the 5-branes form a junction. For non-conformal defects the D3-branes are displaced from the junction point.
We have identified the holographic representation of the defects as probe D3-branes in the $AdS_6\,{\times}\, S^2\,{\times}\,\Sigma$ supergravity duals of the 5d SCFTs and studied conformal and non-conformal defects.

The conformal defects preserve half of the supersymmetries of the ambient SCFT. 
The embeddings wrap an $AdS_4$ in $AdS_6$ and are localized on $S^2$ and on the Riemann surface $\Sigma$.
On $\Sigma$ they are located at points where the $AdS_6$ and $S^2$ radii are extremal.
For the solutions considered here there is exactly one such extremal point on $\Sigma$, and we found one half-BPS conformal defect described by a probe D3-brane. The planar defects can be mapped to $S^3$ defects in the 5d SCFTs on $S^5$ by a conformal transformation, and we obtained the defect contribution to the free energy on $S^5$.
It would be interesting to study these defects from the field theory perspective and e.g.\ reproduce their contribution to the free energy on $S^5$ using supersymmetric localization.
It would also be interesting to consider the limit where the number of D3-branes ending on the 5-brane web becomes large.
Such configurations should be described by $AdS_4$ solutions incorporating the backreaction of the D3-branes on the 5-brane web.
At least certain defects may be described by uplifting $AdS_4$ solutions of 6d gauged supergravity to Type IIB, and it would be interesting to further explore this for the theories studied here. Related discussions based on the Brandhuber/Oz solution of massive Type IIA can be found in  \cite{Dibitetto:2018iar,Penin:2019jlf,Faedo:2020nol,Faedo:2020lyw}.

We also considered non-conformal defects which preserve half of the Poincar\'e supercharges of the ambient SCFT and a $U(1)$ R-symmetry, while breaking the conformal supersymmetries.
The defects are described by D3-brane embeddings which wrap (part of) $AdS_4$ in $AdS_6$ but where the location on $\Sigma$ depends on the $AdS_4$ radial coordinate in Poincare coordinates. 
Fluctuations of the D3-brane away from the conformal point on $\Sigma$ are dual to a pair of relevant operators, and the non-conformal embeddings describe defect RG flows triggered by combinations of these operators.

The features of the non-conformal embeddings suggest a natural brane-web interpretation: 
For special values of the phase of the complex number formed out of the two mass parameters triggering the flow, the embeddings approach, without ever reaching, one of the poles on the boundary of $\Sigma$, which represent the external 5-branes of the associated 5-brane junction.
These embeddings seem to describe D3-branes displaced in the 5-brane web along one of the external 5-branes.
Embeddings for more general values of the phase can be realized if there are two D3-branes ending from opposite sides on the brane web, corresponding to two D3-branes on antipodal points of the $S^2$ in the supergravity solution.
These two embeddings reach a generic point on the boundary of $\Sigma$ at a finite value of the $AdS_4$ radial coordinate. 
Since the $S^2$ collapses on the boundary of $\Sigma$, the two D3-brane embeddings can combine to form a smooth worldvolume without boundary. 
In the brane web these embeddings seem to correspond to more general displacements of the entire D3-brane away from the 5-brane junction, which do not need to be along one of the external 5-branes.

An interesting subject in the context of defect RG flows are RG monotones. It would be interesting to study the contribution of the non-conformal defects to the entanglement entropy, which can be done without computing the backreaction of the branes using the method of \cite{Karch:2014ufa,Chalabi:2020tlw}.

\let\oldaddcontentsline\addcontentsline
\renewcommand{\addcontentsline}[3]{}
\begin{acknowledgments}
The work of M.G.~was  supported, in part,  by the National Science Foundation under grant PHY-19-14412. 
The work of C.F.U.~was supported, in part, by the US Department of Energy under Grant No.~DE-SC0007859 and by the Leinweber Center for Theoretical Physics.
\end{acknowledgments}
\let\addcontentsline\oldaddcontentsline

\appendix
\renewcommand\theequation{\Alph{section}.\arabic{equation}}

\section{D3-brane BPS condition}\label{app:D3-kappa} 

In this appendix we discuss the BPS conditions for the D3-branes introduced in sec.~\ref{sec:D3-embeddings}.
The preserved supersymmetries are generated by Killing spinors $\epsilon$ that satisfy \cite{Bergshoeff:1996tu, Cederwall:1996pv,Cederwall:1996ri}
\begin{align}\label{eq:D3-kappa-0}
 \Gamma_\kappa\epsilon&=\epsilon~.
\end{align}
To evaluate this condition we collect the relevant details on the Killing spinors for the $AdS_6$ solutions.
A general Killing spinor is expanded in a basis of $AdS_6\times S^2$ Killing spinors as
\begin{align}
 \epsilon&=\sum_{\eta_1\eta_2}\chi^{\eta_1\eta_2}\otimes\zeta_{\eta_1\eta_2}~,
 &
 \chi^{\eta_1\eta_2}&=\epsilon_{AdS_6}^{\eta_1}\otimes \epsilon_{S^2}^{\eta_2}~,
\end{align}
where $\eta_1,\eta_2=\pm$. 
One of the Killing spinors, $\chi^{++}_\alpha$, is chosen arbitrarily, and then
\begin{align}\label{eq:chi-rest}
 \chi^{+-}_\alpha&=(\id_8\otimes \gamma_{(2)}) \chi^{++}_\alpha~,
 &
 \chi^{-+}_\alpha&=( \gamma_{(1)}\otimes \id_2) \chi^{++}_\alpha~,
 &
 \chi^{--}_\alpha&=(\gamma_{(2)}\otimes \gamma_{(2)}) \chi^{++}_\alpha~.
\end{align}
The coefficient spinors $\zeta_{\eta_1\eta_2}$ are parameterized as
\begin{align}\label{eq:zeta}
 \zeta_{++}&=\begin{pmatrix} \bar\alpha\\ \beta \end{pmatrix}~,
 &
 \zeta_{--}&=\begin{pmatrix} -\bar\alpha\\ \beta \end{pmatrix}~,
 &
 \zeta_{\eta_1,-}&=i\nu \eta_1 \zeta_{\eta_1,+}~.
\end{align}
Thus,
\begin{align}
 \label{eq:eps}
 \epsilon&=\sum_{\eta}\left(\chi^{\eta,+}+i\nu \eta\chi^{\eta,-}\right)\otimes \zeta_{\eta,+}~,
\end{align}
The explicit form of the $S^2$ and $AdS_6$ Killing spinors was derived in app.~B.1 and B.2 of \cite{DHoker:2016ujz}.
The metric and Killing spinors are, with constant spinors $\epsilon^{\eta_2}_{S^2,0}$, $\epsilon^{\eta_1}_{AdS_6,0}$,
\begin{align}
 ds^2_{S^2}&=d\theta_2^2+\sin^2\!\theta_2\,d\theta_1^2~,
 &
 \epsilon_{S^2}^{\eta_2}&=\exp\left(\frac{i\eta_2}{2}\theta_2\sigma_2\right)\exp\left(-\frac{i}{2}\theta_1\sigma_3\right)\epsilon_{S^2,0}^{\eta_2}~,
\nonumber\\
ds^2_{AdS_6}&=dr^2+e^{2r}dx^\mu dx_\mu~,
 &
 \epsilon_{AdS_6}^{\eta_1}&=e^{\frac{\eta_1}{2}r\gamma_r}\left(1+\frac{1}{2}x^\mu\gamma_\mu\left(\eta_1-\gamma_r\right)\right)\epsilon^{\eta_1}_{AdS_6,0}~.
 \label{eq:AdS6-S2-Killing}
\end{align}

\subsection{Conformal defect}

To find the projection conditions on the Killing spinors it is convenient to start with conformal defects.
For the D3-brane wrapping $AdS_4$, localized at a fixed point in $S^2$ and $\Sigma$, and
using complex notation for the Killing spinors,
\begin{align}
 \Gamma_\kappa\epsilon&=-i\Gamma_0\Gamma_1\Gamma_2\Gamma_3\epsilon
 \nonumber\\
 &=-i(\gamma_r\gamma_0\gamma_1\gamma_2\otimes \id_2\otimes \id_2)\epsilon~.
\end{align}
The condition we aim to implement is
\begin{align}\label{eq:BPS-conf}
 (-i\gamma_{r012}\otimes \id_2)(\epsilon_{AdS_6}^{\eta_1}\otimes \epsilon_{S^2}^{\eta_2})&=\lambda \epsilon_{AdS_6}^{-\eta_1}\otimes \epsilon_{S^2}^{-\eta_2}~,
\end{align}
with $\lambda^2=1$.
Note that $\chi^{-\eta_1,-\eta_2}=(\gamma_{(1)}\otimes \gamma_{(2)})\chi^{\eta_1\eta_2}$.
For $x^3=x^4=0$ the $\gamma_3$, $\gamma_4$ terms drop out in the $AdS_6$ Killing spinor in (\ref{eq:AdS6-S2-Killing}).
Moreover, the matrices multiplying $\epsilon_{S,0}^\pm$ are constant on the brane embedding.
The relation (\ref{eq:BPS-conf}) can be realized by constraining the constant spinor to satisfy
\begin{align}
	\label{eq:eq:BPS-conf-0}
 (-i\gamma_{r012}\otimes \id_2)(\epsilon^{+}_{AdS_6,0}\otimes \epsilon_{S^2,0}^{+})&=\lambda (\gamma_{(1)}\otimes R^{-1}\gamma_{(2)} R)(\epsilon^{+}_{AdS_6,0}\otimes \epsilon_{S^2,0}^+)~,
\end{align}
where $R=\exp\left(\frac{i}{2}\theta_2\sigma_2\right)\exp\left(-\frac{i}{2}\theta_1\sigma_3\right)$.
This is a constant projection condition which squares to one, and implies (\ref{eq:BPS-conf}).
With (\ref{eq:eps}) this leads to 
\begin{align}
	\Gamma_\kappa\epsilon&=\lambda\sum_{\eta} \left(\chi^{-\eta,-}+i\nu\eta\chi^{-\eta,+}\right)\otimes\zeta_{\eta,+}~,
\end{align}
where $\zeta_{++}=(\bar \alpha,\beta)$ and $\zeta_{-+}=i\nu (-\bar\alpha,\beta)$.
The BPS condition becomes
\begin{align}
	\Gamma_\kappa\epsilon-\epsilon \ = \ & (\lambda+1)\left(\chi^{--}-\chi^{++}+i\nu (\chi^{-+}-\chi^{+-}\right)\otimes \begin{pmatrix}\bar\alpha \\ 0\end{pmatrix}
	\nonumber\\
	& + (\lambda-1)\left(\chi^{--}+\chi^{++}+i\nu (\chi^{-+}+\chi^{+-}\right)\otimes \begin{pmatrix}0 \\ \beta\end{pmatrix}~.
\end{align}
The two choices of projectors lead to the BPS conditions
\begin{align}\label{eq:BPS-app-conf}
&&	\lambda&=+1: & \alpha&=0~, &&\nonumber\\
&&	 \lambda&=-1: & \beta&=0~.&&
\end{align}
These conditions can be further evaluated using the expressions for the Killing spinor components in (4.9) of \cite{DHoker:2016ujz}.
This will be covered by the discussion for more general non-conformal embeddings below.
Embeddings at points satisfying (\ref{eq:BPS-app-conf}) preserve half the background supersymmetries,
and realize the sub-superalgebra $C(3)$ of $F(4)$ \cite{Frappat:1996pb} (see also table 1 of \cite{Gutperle:2017nwo}).

\subsection{Non-conformal defect}

Now we consider more general embeddings, where the position on $\Sigma$ is allowed to depend on the radial coordinate $r$ of $AdS_6$, such that the embedding is parametrized by a function $w(r)$.
Then
\begin{align}
	\Gamma_\kappa\epsilon&=\frac{-i}{\sqrt{f_6^2+4\rho^2 |w^\prime|^2}}\left(f_6\Gamma_r+2\rho w^\prime \Gamma_w + 2\rho \bar w^\prime \Gamma_{\bar w}\right)\Gamma_{012}\epsilon~,
\end{align}
where
\begin{align}
	\Gamma_w&=\gamma_{(1)}\otimes\gamma_{(2)}\otimes \gamma_w~,
	&	\gamma_w&=\begin{pmatrix} 0 & 0  \\ 1 & 0 \end{pmatrix}~,
	&
	\Gamma_{\bar w}&=\gamma_{(1)}\otimes\gamma_{(2)}\otimes \gamma_{\bar w}~,
	&\gamma_{\bar w}&=\begin{pmatrix} 0 & 1  \\ 0 & 0 \end{pmatrix}~.
\end{align}
With the Killing spinor (\ref{eq:eps}) and using the projection condition (\ref{eq:eq:BPS-conf-0}) derived for the conformal defect, this leads to
\begin{align}
	\Gamma_\kappa \epsilon=
	\frac{\lambda}{\sqrt{f_6^2+4\rho^2 |w^\prime|^2}}
	\sum_\eta
	\Big[& f_6\left(\chi^{-\eta,-}+i\nu\eta\chi^{-\eta,+}\right)\otimes \zeta_{\eta,+}
	\nonumber\\ &
	-2\rho (\gamma_1\otimes \id_2)\left(\chi^{\eta,+}+i\nu\eta\chi^{\eta,-}\right)\otimes \left(w^\prime \gamma_w+\bar w^\prime \gamma_{\bar w} \right)\zeta_{\eta,+}\Big].
\end{align}
Now we impose the following condition
\begin{align}
	\gamma_r \epsilon_{AdS_6,0}^+&= \epsilon_{AdS_6,0}^+
	& &\Rightarrow
	&
	\gamma_r \epsilon_{AdS_6,0}^-&= -\epsilon_{AdS_6,0}^-\,.
\end{align}
Note that $\gamma_r$ commutes with $\gamma_{r012}\gamma_{(1)}$, so the projection condition on $\epsilon_{AdS_6,0}^+$ is compatible with (\ref{eq:eq:BPS-conf-0}).
The condition on $\epsilon_{AdS_6,0}^-$ follows from $\epsilon_{AdS_6,0}^-=\gamma_{(1)}\epsilon_{AdS_6,0}^+$.
In view of the explicit form of the $AdS_6$ Killing spinors, this condition implies
\begin{align}
	\epsilon_{AdS_6}^\pm&=e^{\frac{1}{2}r}\epsilon_{AdS_6,0}^\pm & 
	&	\Rightarrow &
	\gamma_r \epsilon_{AdS_6}^\eta&=\eta\epsilon_{AdS_6}^\eta~.
\end{align}
With this relation 
\begin{align}
	\Gamma_\kappa \epsilon=
	\frac{-i\lambda}{\sqrt{f_6^2+4\rho^2 |w^\prime|^2}}
	\sum_\eta
	\Big[& f_6\left(\chi^{-\eta,-}+i\nu\eta\chi^{-\eta,+}\right)\otimes \zeta_{\eta,+}
	\nonumber\\ &
	-2\rho \left(\eta\chi^{\eta,+}+i\nu\chi^{\eta,-}\right)\otimes \left(w^\prime \gamma_w+\bar w^\prime \gamma_{\bar w} \right)\zeta_{\eta,+}\Big].
\end{align}
Now we can spell out the components on $\Sigma$, using $\zeta_{-+}=i\nu\zeta_{--}$ and the explicit expressions in (\ref{eq:zeta}). This leads to
\begin{align}
	\Gamma_\kappa\epsilon&=
	\frac{\lambda}{\sqrt{f_6^2+4\rho^2 |w^\prime|^2}}
	\begin{pmatrix}
		\left(\chi^{--}-\chi^{++}+i\nu(\chi^{-+}-\chi^{+-})\right)\left(f_6 \bar\alpha +2\rho \bar w^\prime \beta\right)
		\\[1mm]
		\left(\chi^{--}+\chi^{++}+i\nu(\chi^{-+}+\chi^{+-})\right)\left( f_6 \beta -2\rho w^\prime \bar\alpha\right)
	\end{pmatrix}.
\end{align}
The spinor itself reads
\begin{align}
	\epsilon&=
	\begin{pmatrix}
	\left(\chi^{++}-\chi^{--}+i\nu(\chi^{+-}-\chi^{-+})\right)\bar\alpha
	\\[1mm]
	\left(\chi^{++}+\chi^{--}+i\nu(\chi^{+-}+\chi^{-+})\right)\beta
	\end{pmatrix}.
\end{align}
From $\Gamma_\kappa\epsilon=\epsilon$ we thus conclude that the BPS conditions are
\begin{align}\label{eq:BPS-mass-flat-0}
\lambda	\frac{ f_6 \bar\alpha+2\rho\bar w^\prime \beta}{\sqrt{f_6^2+4\rho^2 |w^\prime|^2}}&=-\bar\alpha~,
&
\lambda	\frac{ f_6 \beta-2\rho w^\prime \bar\alpha}{\sqrt{f_6^2+4\rho^2 |w^\prime|^2}}&=\beta~.
\end{align}
These are two complex equations for one complex function.

Eliminating the square root between the two equations, by multiplying the complex conjugate of the first equation by $\beta$ and adding it to the second equation multiplied by $\alpha$, leads to
\begin{align}\label{eq:BPS-mass-flat-1}
   f_6 \alpha\beta + \rho w^\prime \left(|\beta|^2-|\alpha|^2\right)&=0~.
\end{align}
Solving for $w^\prime$ and using the result in (\ref{eq:BPS-mass-flat-0}) uniformizes the square root, 
and leads to the constraint 
\begin{align}
\lambda \sign(|\beta|^2-|\alpha|^2)=1~.
\end{align}
Note that $|\beta|^2-|\alpha|^2=3\nu f_2$, and $f_2$ is non-zero throughout $\Sigma$ for regular solutions, so this always leads to a consistent choice for $\lambda$.

Now to further evaluating (\ref{eq:BPS-mass-flat-1}).
Using the expressions for the spinor components in (4.9) of \cite{DHoker:2016ujz} along with the expression for $B$ leads to
\begin{align}\label{eq:alphabeta}
	\rho^2\bar\alpha^2\bar\beta^2&=\frac{(\partial_w\cG)^2}{6\cG T} = \frac{(\partial_w\cG)^2}{f_6^4}~.
\end{align}
To use this expressions one has to square (\ref{eq:BPS-mass-flat-1}), leading to
\begin{align}
{w^\prime}^2&=\frac{f_6^2\alpha^2\beta^2}{9\rho^2 f_2^2} = \frac{(\partial_{\bar w}\cG)^2}{9f_6^2\rho^4 f_2^2}= \frac{(\partial_{\bar w}\cG)^2}{\kappa^4}~.
\end{align}
For the last equation the explicit expressions for the metric functions were used.
We thus find
\begin{align}\label{eq:BPS-mass-flat}
	\kappa^2 w^\prime &= \tau\partial_{\bar w}\cG~,
\end{align}
with $\tau^2=1$.

We have yet to check the equation of motion resulting from  the action in (\ref{d3act}).
Using the expressions for the metric functions one may write the Lagrangian as
\begin{align}
	L_{\rm D3}&=6e^{3r}\cG T \sqrt{1+\frac{2\kappa^2}{3\cG}|w^\prime|^2}~.
\end{align}
On the BPS configurations the square root evaluates to $T$.
Thus, upon performing the variation first and substituting $w^\prime$ then,
\begin{align}
	\frac{\delta L_{\rm D3}}{\delta \bar w^\prime}&=2\tau e^{3r}\partial_{\bar w}\cG~,
&
	\frac{\delta L_{\rm D3}}{\delta \bar w}&
	=4e^{3r}\partial_{\bar w}\cG + 2e^{3r}\frac{\partial_w\cG \partial_{\bar w}^2\cG}{\kappa^2}~.
\end{align}
From the first equation one concludes
\begin{align}
	\partial_r 	\frac{\delta L_{\rm D3}}{\delta \bar w^\prime}&
	=2e^{3r}(3\tau-1)\partial_{\bar w}\cG+2e^{3r}\frac{\partial_{\bar w}^2\cG\partial_w\cG}{\kappa^2}~.
\end{align}
The equation of motion implies
\begin{align}
	\partial_r 	\frac{\delta L_{\rm D3}}{\delta \bar w^\prime} = \frac{\delta L_{\rm D3}}{\delta \bar w}
&&\Rightarrow & &
	\tau&=1~.
\end{align}
Embeddings satisfying (\ref{eq:BPS-mass-flat}) with $\tau=1$ thus solve the equation of motion and preserve a quarter of the background supersymmetries.

\bibliography{defect}
\end{document}